\RequirePackage[2020-02-02]{latexrelease}
\documentclass[%
superscriptaddress,
 amsmath,amssymb,
 aps,
prd,
twocolumn
]{revtex4}

\usepackage{graphicx}
\usepackage{dcolumn}
\usepackage{bm}
\usepackage{hyperref}
\usepackage[shortlabels]{enumitem}

\usepackage{changes}
\definechangesauthor[name={Lock}, color=red]{LY}
\definecolor{darkgreen}{rgb}{0.0, 0.6, 0.0}
\definechangesauthor[name={VL}, color=darkgreen]{VL}

\newcommand{\R}{{\mathbb R}}

\allowdisplaybreaks 
\begin{document}


\title{Inferring origin-destination distribution of agent transfer in a complex network using deep gated recurrent units}
\author{Vee-Liem Saw}
\email{Vee-Liem@ntu.edu.sg}
\affiliation{Division of Physics and Applied Physics, School of Physical and Mathematical Sciences, Nanyang Technological University, Singapore}
\author{Luca Vismara}
\email{vism0001@e.ntu.edu.sg}
\affiliation{Division of Physics and Applied Physics, School of Physical and Mathematical Sciences, Nanyang Technological University, Singapore}
\author{Suryadi}
\email{sury0013@e.ntu.edu.sg}
\affiliation{Division of Physics and Applied Physics, School of Physical and Mathematical Sciences, Nanyang Technological University, Singapore}
\author{Bo Yang}
\email{yang.bo@ntu.edu.sg}
\affiliation{Division of Physics and Applied Physics, School of Physical and Mathematical Sciences, Nanyang Technological University, Singapore}
\author{Mikael Johansson}
\email{mikaelj@kth.se}
\affiliation{School of Electrical Engineering, KTH Royal Institute of Technology, Stockholm, Sweden}
\author{Lock Yue Chew}
\email{lockyue@ntu.edu.sg}
\affiliation{Division of Physics and Applied Physics, School of Physical and Mathematical Sciences, Nanyang Technological University, Singapore}

%

\date{\today}

\begin{abstract}
Predicting the origin-destination (OD) probability distribution of agent transfer is an important problem for managing complex systems. However, prediction accuracy of associated statistical estimators suffer from underdetermination. While specific techniques have been proposed to overcome this deficiency, there still lacks a general approach. Here, we propose a deep neural network framework with gated recurrent units (DNNGRU) to address this gap. Our DNNGRU is \emph{network-free},  as it is trained by supervised learning with time-series data on the volume of agents passing through edges. We use it to investigate how network topologies affect OD prediction accuracy, where performance enhancement is observed to depend on the degree of overlap between paths taken by different ODs. By comparing against methods that give exact results, we demonstrate the near-optimal performance of our DNNGRU, which we found to consistently outperform existing methods and alternative neural network architectures, under diverse data generation scenarios.


\end{abstract}

\maketitle


Deciphering the origin-destination (OD) pair has been at the heart of various protocols that aim to evaluate the traffic demand and flow within complex systems. Interest in OD pairs results from the basic information it encodes on the distribution of people, materials, or diseases which has direct bearing on the socioeconomic phenomena of human mobility, resource allocation, and epidemic spreading. An intrinsic utility in gaining knowledge of the OD distribution is that paths with higher transfer rates can be enhanced to improve system's efficiency, or blocked to impede the transfer of malicious/undesirable entities. By far the most intensively studied OD problem for a complex system is the estimation of OD traffic of an Internet network from measurable traffic at router interfaces \cite{Coates02}. Collection of link traffic statistics at routers within a network is often a much simpler task than direct measurements of OD traffic. The collected statistics provide key inputs to any routing algorithm, via link weights of the open shortest path first (OSPF) routing protocol. Shortly after, similar techniques have been adapted for the problem of identifying transportation OD by measuring the number of vehicles on roads \cite{Krui37,Tebaldi98,Bera11,Dey20}. Such inference of vehicular OD has been used by Dey et al. \cite{Dey20} to give better estimates of commuters' travel time, or by Saberi et al. \cite{Saberi17} to understand the underlying dynamical processes in travel demand which evolve according to interactions and activities occurring within cities.

Several techniques have been developed in an effort to estimate OD information from link/edge counts. These include expectation-maximisation \cite{Vardi96}, entropy maximisation (or information minimisation) \cite{VZ78,Willumsen78,VZ80}, Bayesian inference \cite{Dey94,Tebaldi98,Hazelton00,Carvalho14}, quasi-dynamic estimations \cite{Cascetta13,Bauer18}, and the gravity model \cite{Balcan09,Dragu19,Ciavarella21}. As the number of OD degrees of freedom (quadratically proportional to number of nodes) generally outnumbers the number of link counts (linearly proportional to number of nodes), the major issue concerning OD estimation is that the problem is severely underdetermined. Vardi \cite{Vardi96} attempted to resolve this issue by treating the measurements of OD intensities as Poisson random variables, and used expectation-maximisation with moments to figure out the most likely OD intensities giving rise to the observed measurements of link counts. The non-unique solutions due to underdetermination had also been addressed through the principle of entropy maximisation, as well as by Bayesian inference through the assumption of prior OD matrices. Alternatively, ambiguities in OD inferences were treated by quasi-dynamic method using prior knowledge about historical trip data, or through parameters calibration of the gravity model based on zonal data. Invariably, these methods lead to large uncertainties and unreliable OD inferences.

In this work, we introduce a new OD inference approach using deep neural network (DNN) and a method based on linear regression (LR). We regress for the \emph{probabilities} $\zeta_{ij}$ of an agent going from origin $i$ to destination $j$, instead of predicting the \emph{actual number of agents} in the OD matrix. These quantities $\zeta_{ij}$ (also known as ``fan-outs'' \cite{Gunnar04}) are assumed to be stationary throughout the period of interest, and are thus always the same. For example, in a bus system, commuters in the morning have some preferred $\zeta_{ij}$, so the fan-outs are constant. But the number of people arriving the bus stops or boarding the buses need not be the same each time. This happens when a next bus arrives relatively quickly after the previous bus has left, with fewer people boarding it. Consequently, simple linear regression can be implemented to obtain $\zeta_{ij}$ from repeated measurements over the period. The actual OD numbers are just the total numbers from the origins (which are easily measurable) multiplied by $\zeta_{ij}$. This approach overcomes the weakness of underdetermined system in previous works, where repeated measurements would correspond to different OD intensities, which cannot be combined. As a result, improvement in prediction accuracy is attained over earlier approaches such as expectation-maximisation and Bayesian inference, which we will demonstrate on a common network in Section \ref{comparemethods}.

Next, we go beyond linear regression by training a DNN with supervised learning to predict the OD probabilities $\zeta_{ij}$. Our approach is thus applicable to any arbitrary network that connects between a set of origin nodes and a set of destination nodes. In other words, our framework is \emph{network-free}, i.e. directly applicable to any network without requiring explicit modelling and analysis. Furthermore, the DNN is composed of gated recurrent units (GRU) \cite{GRU} which capture and process the temporal information in our data. The use of GRU is also necessary because our input data is in the form of a time-series, and the same GRU architecture can process time-series input data of arbitrary length. (In contrast, densely connected feedforward DNN would have the number of input nodes dependent on the length of the time-series data.) In comparison, analytical statistical frameworks like the Vardi's algorithm \cite{Vardi96} and the Bayesian methods \cite{Tebaldi98} necessitate explicit encoding of the network (referred to as the ``routing matrix'', related to the adjacency matrix of a graph) before implementation. Temporal information is not exploited as each datum is treated as being independent from the others.

We harness this DNN framework to study general complex networks with different topologies like lattice, random, and small-world, as well as real-world networks,  to glean how network topology affects the accuracy in predicting the OD probabilities. Recently, there is great interest in using deep learning methods to solve problems in applications modelled by complex network, such as predicting the dismantling of complex systems \cite{Grassia21}, and on contagion dynamics \cite{Murphy21}. While these papers trained deep learning approaches to identify topological patterns on dynamical processes in networks, they have not explored how topology of different complex networks affect the OD prediction accuracies. Incidentally, Ref. \cite{Murphy21} made use of OD information of human mobility to study the spread of COVID-19 in Spain through a complex network model. They compared their DNN approach with a maximum likelihood estimation (MLE) technique and found that the former outperforms the latter. This outcome is analogous to our case because Vardi's expectation-maximisation algorithm with moments is in fact an MLE method. Moreover, unlike Refs.\ \cite{Grassia21,Murphy21} which implemented graph neural network and graph attention network that made use of convolution of neighbouring nodes to significantly improve performance, we design a GRU architecture which leverages on temporal information to predict OD probabilities.

\section{Results}

\subsection{Data measurements}

Before presenting the linear regression (LR) and deep neural network (DNN) approaches to infer the origin-destination distribution, we elaborate on the data types that are easily measurable within generic real-world complex networks. In our setup, we primarily consider the dataset to be a time-series of $T$ time steps where the number of agents are tracked at every single time step. Then, at a single time step, the measurable quantities are: $x_i$ which is the number of agents leaving origin $i$, $y_j$ which is the number of agents arriving at destination $j$, as well as $F_{ab}$ which is the number of agents on the directed edge from node $a$ to node $b$ of the complex network.

We employ $x_i$ and $y_j$ for LR due to its formulation. For general complex network, LR can only serve as an approximate estimator, unless the network is relatively simple enough to allow for an exact representation. In contrast, conventional algorithms for origin-destination estimation \cite{Vardi96,Tebaldi98} use the number of agents on the edges, $F_{ab}$. Hence, in developing our DNN framework, we will be using only $F_{ab}$ but not including $x_i$ and $y_j$. This mode of data measurement is used in Sections \ref{GCN} and \ref{scenarios}.

Nevertheless, in order for us to provide a fair evaluation of our DNN framework as compared to Vardi's \cite{Vardi96} and Tebaldi-West's \cite{Tebaldi98} approaches, we will consider a separate type of measurement in our comparison against these traditional approaches. In this case, we count the number of agents in $x_i$ and $F_{ab}$ over some fixed time interval, i.e. these numbers are aggregated instead of the actual numbers at every time step. Then, $T$ such measurements are collected (where this $T$ is now the number of independent aggregated measurements, instead of the length of the time series). This means of aggregated measurement is in fact implemented by Vardi and Tebaldi-West, although Vardi's formulation uses two variables: one counts the number of agents from origin to destination, and the other counts the agents on the edges, which he denotes by $X_i$ and $\vec{Y}$ respectively. To avoid confusion with our notation, we have renamed our $x_i$ as $V_i$ while retaining the use of the Vardi's vector of aggregated number of agents on all edges $\vec{Y}$ in our formulation. We will use these aggregated variables in our DNNGRU studies as well as that of our LR formulated using these variables described in the Supplemental Material (SM). Note that this mode of data measurement is employed in Section \ref{comparemethods}.

\subsection{A Linear Regression Approach}

\begin{figure*}
\centering
\includegraphics[width=17cm]{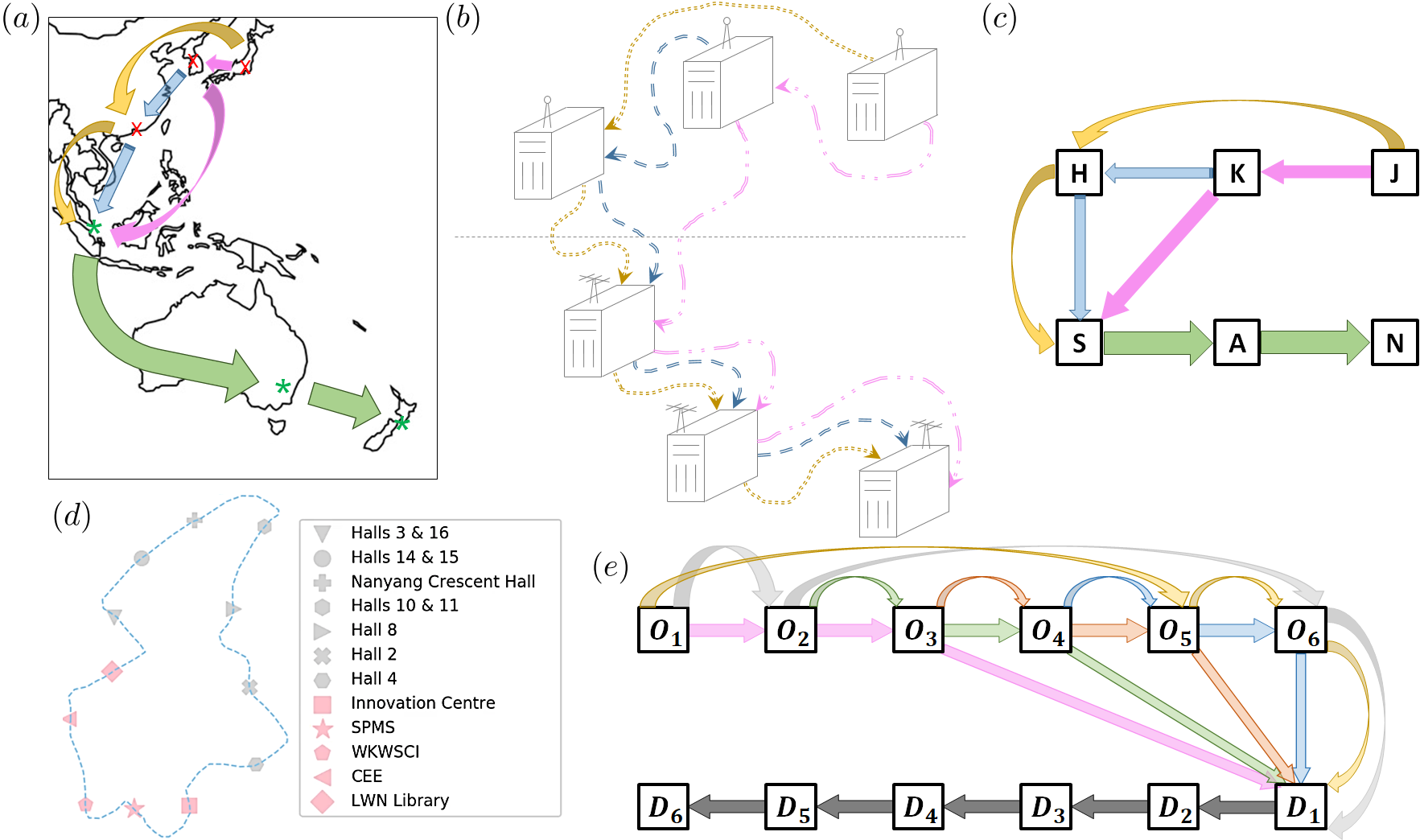}
\caption{a: Hypothetical example of three origin data servers (red x) connected to three destination data servers (green $*$) in Asia-Pacific. b: Pictorial visualisation of data server connections of (a). c: Network representation of (a, b). d: Blue Route of the NTU campus shuttle bus service \cite{Vee2019,Quek2020}. e: Network representation of (d), where the nodes are linked via a \emph{semi-express} configuration \cite{Aramsiv21,Vee2021}. Each of the six coloured arrows represents one semi-express bus picking up commuters from distinct subsets of origins. Subsequently, all buses allow alighting at all destinations $D_1,\cdots, D_6$.}
\label{fig1}
\end{figure*}

Let us consider a complex network of nodes where agents from an origin node can end up at any destination node in the network. In our context, an edge in the network corresponds to a carrier route that facilitates transfer of agents between the two nodes connected by it. For example, data server hubs are linked through a series of fibre-optic cables (\emph{carriers}). As not all data server hubs are directly connected due to geography, data (\emph{agents}) transfer between a pair of data servers may traverse other data servers along the way, as depicted in Fig.\ \ref{fig1}(a, b, c). In another example, the bus service in Fig.\ \ref{fig1}(d) comprises a loop of $12$ bus stops served by buses going around. In other words, buses (\emph{carriers}) must sequentially traverse one bus stop after another to deliver commuters (\emph{agents}) in some fixed order along the prescribed loop.

In a network with $M_O$ origins and $M_D$ destinations, there are $M_O(M_D-1)$ free parameters which quantify the probability distribution of agent-transfer from one node to another. This is because at each of the $M_O$ origins, there are $M_D$ destinations and these probabilities sum to $1$. Whilst we do not know where each agent goes, we can write down this general multivariate linear system:
\begin{align}\label{generalmultivariate}
y_j=\sum_{i=1}^{M_O}{\zeta_{ij}x_i},\textrm{ for }j=1,\cdots,M_D,
\end{align}
subject to the constraints $\displaystyle\sum_{j=1}^{M_D}{\zeta_{ij}}=1$, for $i=1,\cdots,M_O$. Here, $x_i$ are the number of agents the carrier picks up from origins  $i=1,\cdots,M_O$ whilst $y_j$ are the number of agents that carrier delivers at destinations $j=1,\cdots,M_D$. The sought after quantities are $\zeta_{ij}$, denoting the OD probabilities of agents from $i$ to $j$. As Eq.\ (\ref{generalmultivariate}) does not account for agent-transfer through the edges with different travelling paths between the same origin and destination, it serves basically as an approximation model.

Assuming that $\zeta_{ij}$ are stationary and hence independent of time, repeated measurements will yield different $x_i$ and $y_j$ leading to an overdetermined set of equations given through Eq.\ (\ref{generalmultivariate}). The OD coefficients $\zeta_{ij}$ can then be deduced by linear regression (LR), given dataset $(x_i,y_j)$ to be fitted \cite{LR12}. In other words, we analytically solve for $\zeta_{ij}$ through the minimisation of the mean squared error. As LR gives coefficients $\zeta_{ij}\in\R$, there are occasions when $\zeta_{ij}<0$. We handle this by minimally shifting the entire $\zeta_{ij}$ to make all $\zeta_{ij}\geq0$, and then normalise them so that the $M_O$ constraints below Eq.\ (\ref{generalmultivariate}) are satisfied. This formulation constitutes a linear framework of OD estimation modelled by a complex network that relates to a particular system of interest. Depending on the system under examination, the variables $x_i$ and $y_j$ are determined from the relevant measured empirical data. We have compared our LR approach with quadratic programming where the optimisation is performed by imposing the additional constraint that $\zeta_{ij}$ is non-negative. While quadratic programming is observed to perform consistently better than LR at small data size, LR takes a significantly shorter time to reach the solution relative to quadratic programming. Both approaches nonetheless converge to the same solution when sample size increases.

In the case of the bus system, publicly accessible information such as the number of people on buses and the duration buses spend at bus stops \cite{busurl} can be used to provide $x_i = $ number of people boarding and $y_j = $ number of people alighting the associated bus. In the context of data servers, $x_i$ and $y_j$ are deducible from the traffic load passing through the fibre-optic cables carrying packet bits since increased/decreased load is due to $x_i$/$y_j$ from/to a data server.

\subsection{DNN with Supervised Learning}
\begin{figure*}
\centering
\includegraphics[width=17cm]{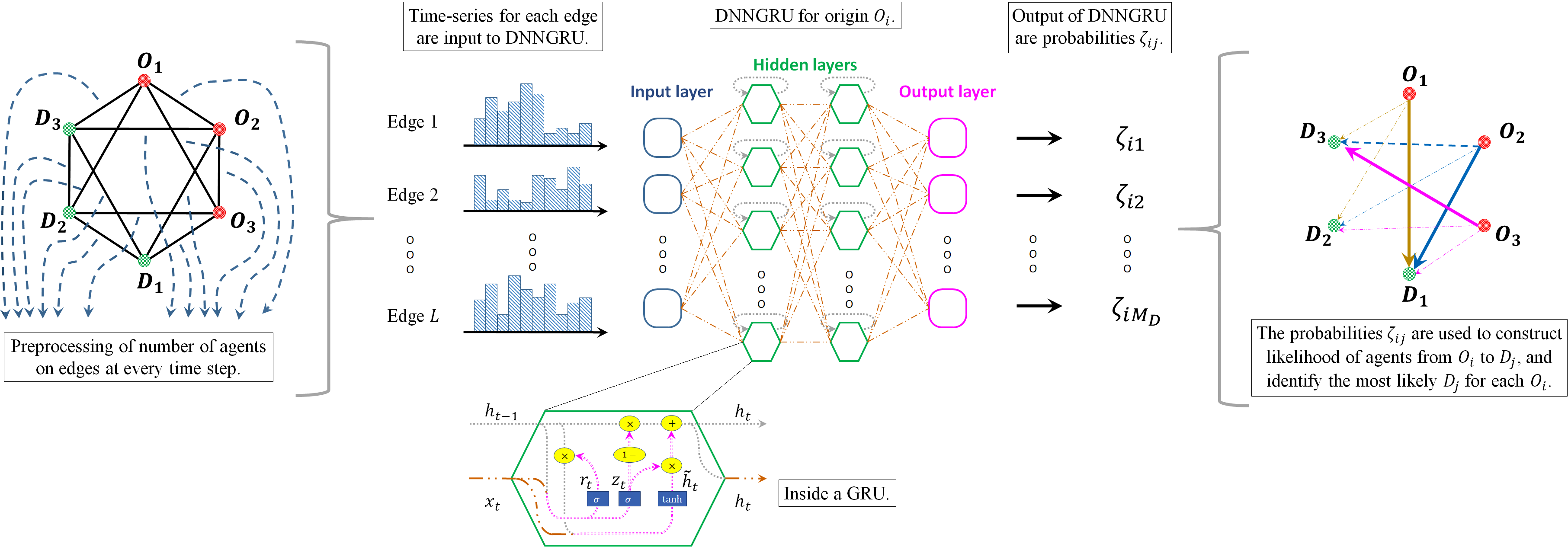}
\caption{Number of agents from each edge of a complex network are measured. After preprocessing them into time-series of length $T$ for every edge, they are fed into a DNNGRU and trained by supervised learning to predict $\zeta_{ij}$. Separate DNNGRUs are trained for each origin $O_i$ and each time-series length $T$. From the outputs $\zeta_{ij}$, we construct the likelihood of agents going from $O_i$ to $D_j$, as well as identify the most likely destination from each origin node.}
\label{fig2}
\end{figure*}

The main idea of our approach is to determine the OD probabilities $\zeta_{ij}$ from easily accessible or directly measurable information, specifically the number of agents on the edges of the complex network. This ability to infer $\zeta_{ij}$ is particularly important when it is impossible to measure $\zeta_{ij}$. For example, it is extremely challenging for investigators to figure out the tracks of nefarious hackers/fugitives who would doubtlessly obfuscate their direct transfer of data packets across various data servers. Such problems like commuters in transportation systems as well as fugitive hunting motivate the inference of the most likely destination of agent transfer, from an origin, so that we can better improve service and connectivity or to better locate a target's whereabouts. We will thus measure the prediction accuracy of our algorithms developed in this paper in predicting the most popular destination with respect to an origin.

In order to deal with these more general situations, we develop a DNN architecture composed of GRU (i.e. DNNGRU) to output $\zeta_{ij}$ by supervised learning, where training may be performed on real-world datasets which are amply available. In addition, labelled data can also be generated through simulation from systems with known characteristics for DNN to generalise from them to new unseen inputs. Fig.\ \ref{fig2} gives an overview on the approach of this work, where information of the number of agents on the edges (e.g. number of people on buses after leaving bus stops) are passed into DNNGRU to infer OD probabilities of the network.

Our DNN can take as input a more generic form of the dataset with $(x_i,y_j)$ implicitly contained, as compared to LR described in the previous subsection. It thus encompasses a more general data and model structure of the system-under-study such that the actual mechanism of the carriers may be complex and nonlinear. Supervised learning then allows our DNN to learn directly and more generally from labelled $\zeta_{ij}$. Furthermore, we adopt a DNN with sigmoid activation to naturally output the correct range of $\zeta_{ij}\in[0,1]$. An important characteristic of our designed DNN architecture is the incorporation of GRU to enable inference from time series.
The GRU is a recurrent unit with hidden memory cell that allows for information from earlier data to be combined with subsequent data in the time-series. This is crucial as the dataset $(x_i,y_j)$ may carry temporal information. For instance, buses recently picking up commuters would leave less people for subsequent buses, hence cascading effects (like bus bunching, overtaking, load-sharing \cite{Vee2019,Chew2021}) induce deviations from Eq.\ (\ref{generalmultivariate}) which ignores temporal correlation as it treats each datum independently.


\subsection{Comparison amongst DNNGRU, LR, EM and other traditional methods} \label{comparemethods}

\begin{figure*}
\centering
\includegraphics[width=17cm]{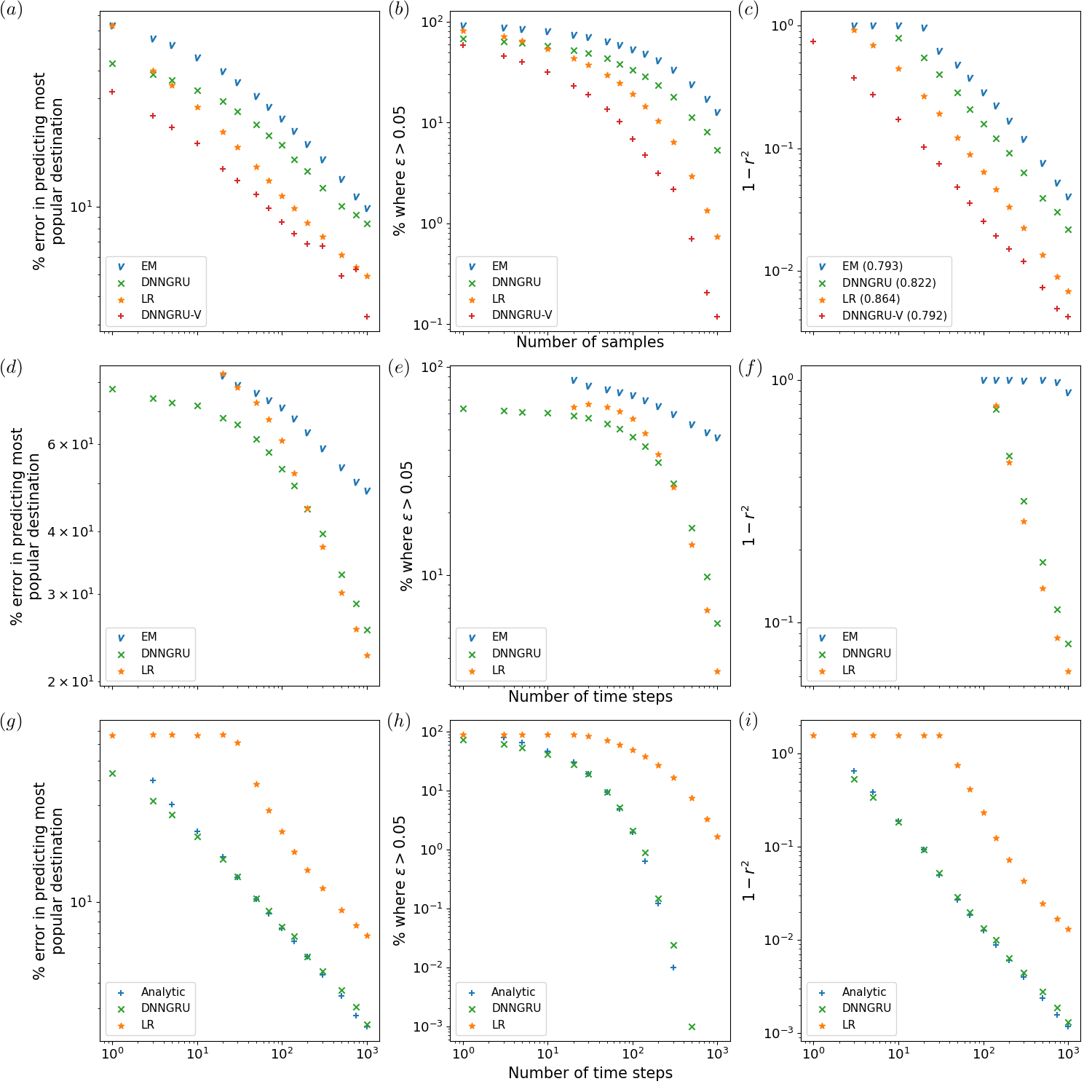}
\caption{a, b, c: Comparison amongst DNNGRU and Vardi's EM algorithm for Vardi's network \cite{Vardi96} as well as LR and DNNGRU-V with additional data on the total number of agents from the origin nodes. Shown in parentheses of the legend in (c) are the exponents of a power law fit. d, e, f: Corresponding comparison amongst DNNGRU, LR, and Vardi's EM algorithm \cite{Vardi96} for the loop in Fig.\ \ref{fig4}(a). g, h, i: Corresponding comparison amongst DNNGRU, LR and an analytical averaging for a lattice with $M_O=M_D=3$ in Fig.\ \ref{fig4}(g). Here, LR in Eq. (\ref{generalmultivariate}) is only an approximation as it does not track the exact number of $y_j$. In (h), note that for $T>500$, DNNGRU and the analytic treatment have $0\%$ error in predicting $\zeta_{ij}$ to be within $0.05$ from the true value. Therefore, this would be $-\infty$ on the log-scale, which is not shown.}
\label{fig3}
\end{figure*}

We directly compare our DNNGRU and LR approaches with the traditional Vardi's expectation-maximisation (EM) algorithm, Willumsen's entropy maximisation, as well as Bayesian method on Vardi's network as a testbed \cite{Vardi96}. For this purpose, we adapt the generation of data according to the approach of Vardi: The number of agents on the edges are measured over an extended time period \cite{Vardi96,Tebaldi98}. This would be a setup where a counter tracks the overall number of agents that passes through an edge throughout the entire day, for instance. In this case, samples from various days are independent, with the number of people $X_k$ generated for each origin-destination pair being a Poisson random variable with a fixed mean $\lambda_k$. Vardi's network has four nodes, and consequently 12 OD components in $\vec{X}$ (see Methods for a review of Vardi's EM algorithm and notations).

We generate datasets in the following manner: For each dataset, each $\lambda_k$ is an integer chosen from $[1, 20]$. Then, an $\vec{X}$ is drawn, with $\vec{Y}$ computed via Eq.\ (\ref{YAX}). Here, $\vec{Y}$ is the vector of number of agents on the edges of the network. The $\zeta_{ij}$ are computed from appropriately normalising $\vec{\lambda}$ with respect to each origin node. In other words, the ground truth of $\zeta_{ij}$ are: $\zeta_{11}=\lambda_1/(\lambda_1+\lambda_2+\lambda_3)$, etc. Therefore, for each dataset, we have $\zeta_{ij}$ and $\vec{Y}$. The goal is to infer $\zeta_{ij}$ from $\vec{Y}$. We collect $T$ samples of $\vec{Y}$ in each dataset for that specific $\zeta_{ij}$, and this provides the input for DNNGRU, as well as Vardi's algorithm. More datasets can be generated by resetting $\vec{\lambda}$. For LR to be applicable as an exact model, we need to input an additional piece of information, viz. the total number of people originating from each origin node, $\vec{V}$. The exact solution of $\zeta_{ij}$ by LR is given in SM. This additional information is also provided to train an alternative DNNGRU, which we call ``DNNGRU-V'' to distinguish it from the version that does not include $\vec{V}$ as its input.

Separate DNNGRUs (as well as DNNGRU-Vs) are trained to output $\zeta_{ij}$ for each $i$, so a network with $M_O$ origin nodes has $M_O$ DNNGRUs. Furthermore for each $i$, we let $T=1, 3, 5, 10, 20, 30, 50, 70, 100, 140, 200, 300, 500, 750, 1000$, and apply transfer learning (see Methods) to separately train different DNNGRUs for different $T$. Training comprises 250k datasets (where $\zeta_{ij}$ has been randomised), with another 10k as validation during training. Additionally, 50k datasets (with randomised $\zeta_{ij}$) are for testing.

After obtaining these probabilities $\zeta_{ij}$, we determine the most popular destination $j$ with respect to origin $i$ by selecting the corresponding $j$ with the highest probability $\zeta_{ij}$. We also determine how accurate is the prediction of $\zeta_{ij}$ with respect to the true value, by two different measures, viz. the difference $\varepsilon =|\zeta_{ij, predicted} - \zeta_{ij, true}|$,  as well as the coefficient of determination $r^2$ (see Method) for the straight line fit of $\zeta_{ij,predicted}=\zeta_{ij,true}$. Fig.\ \ref{fig3}(a) shows the percentage error $E$ in predicting the most popular destination, versus the number of samples $T$. Fig.\ \ref{fig3}(b) shows the percentage of the predictions where $\varepsilon>0.05$, versus $T$. Fig.\ \ref{fig3}(c) shows $1-r^2$ for the straight line fit of $\zeta_{ij, predicted}=\zeta_{ij, true}$, versus $T$.

Evidently, DNNGRU outperforms EM with the usual data of number of agents on all the edges. We also observe LR to perform better than DNNGRU and EM, which results from the exploitation of the additional data of $\vec{V}$. Note that $\vec{V}$ is easily obtainable for example by placing a counter that tracks every agent that leaves that node, without actually knowing where the agents are going. The use of $\vec{V}$ has also resulted in improvement in the outcome of DNNGRU-V, where it is observed to have superior performance relative to LR. DNNGRU (and DNNGRU-V) is advantageous here by benefiting from the many weights and biases to be tuned from supervised learning with a deep neural network. Basically, the enhanced performances of DNNGRU and DNNGRU-V over Vardi's algorithm and LR, respectively, arise from the gains accrued by learning from data. Nonetheless, the performance of DNNGRU-V would converge to that of LR in the asymptotic limit of large $T$ since LR gives exact solution (see Fig.\ 1(a) SM). On the other hand, LR shows better performance compared to EM due to its lower modelling uncertainties from fewer underlying assumptions. As for entropy maximisation and Bayesian inference, the results are below par and hence not displayed. The latter two methods have not addressed the underdetermination problem and consequently are still burdened by the high uncertainty of their results. See SM for more details.

Incidentally, results corresponding to Fig.\ \ref{fig3}(c) where the predictions are the actual OD intensities instead of $\zeta_{ij}$ are shown in Fig.\ 1(b) of SM. To predict the actual OD intensities, we just multiply the predicted probabilities $\zeta_{ij}$ with the corresponding average total number of people at the origins. This is fully in accordance to how Vardi intended to apply his expectation-maximisation algorithm. In terms of the actual OD intensities, again LR, DNNGRU, DNNGRU-V all outperform EM.

A performance analysis and comparison between DNNGRU (without $\vec{V}$), LR, and EM have also been performed for a loop network (Fig.\ \ref{fig4}(a)) with the same outcome observed. Details are given in Methods, with corresponding results displayed in Figs.\ \ref{fig3}(d, e, f). The data generation here is not based on measuring the agents over a specified time interval, as prescribed by Vardi's network \cite{Vardi96}. Instead, it is based on the realistic flow of agents from one edge to the next edge, which we elaborate further in the next subsection on general complex networks. The point here is, DNNGRU and LR are superior to Vardi's EM algorithm. All three methods receive the same information of the number of agents on the edges at every time step, over $T$ time steps.

\subsection{General complex networks}\label{GCN}
\begin{figure*}
\centering
\includegraphics[width=17cm]{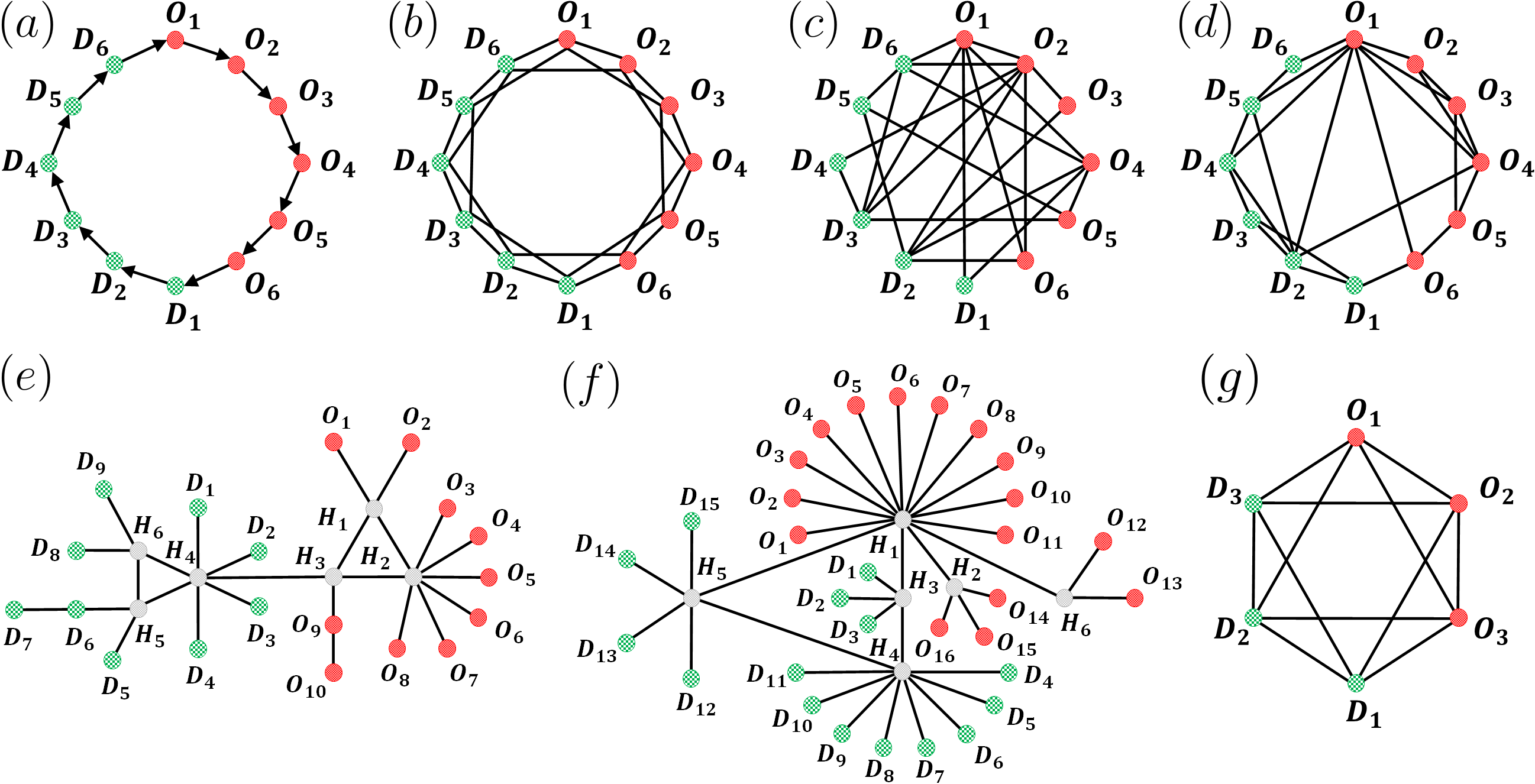}
\caption{a: Directed loop, with 11 ADE. b: Lattice, with 38 ADE. c: Random, with 32 ADE, d: Small-world, with 24 ADE, e: VinaREN, with 24 ADE, f: MYREN, with 40 ADE, g: Smaller lattice, with 18 ADE. In each network, $O_i$ (red) are origins, $D_j$ (green) are destinations. (a-d) have 6 origins and 6 destinations, whilst (e) has 10 origins and 9 destinations with 6 intermediate nodes (grey), (f) has 16 origins and 15 destinations with 6 intermediate nodes (grey). The smaller lattice in (g) has 3 origins and 3 destinations. (b-d) all have 24 edges (48 directed edges, since each edge represents both directions), so they differ on their topologies being a lattice, random, or small-world network.}
\label{fig4}
\end{figure*}

\begin{figure*}
\centering
\includegraphics[width=17cm]{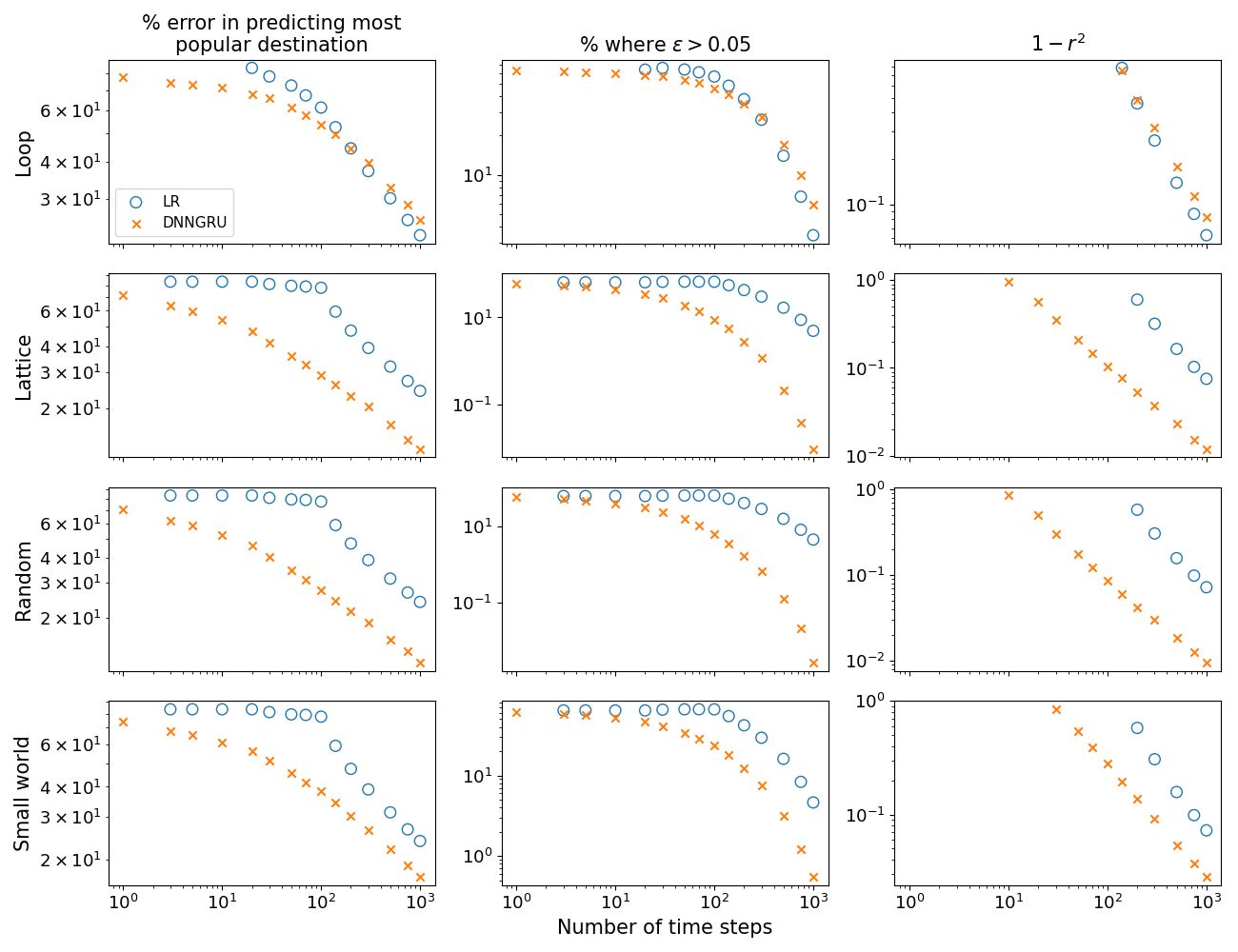}
\caption{DNNGRU outperforms LR in the lattice, small-world and random networks with the same number of nodes and edges, as LR serves as an approximation. For the loop, LR is an exact model and asymptotically matches the performance of DNNGRU.}
\label{fig5}
\end{figure*}
\begin{figure*}
\centering
\includegraphics[width=17cm]{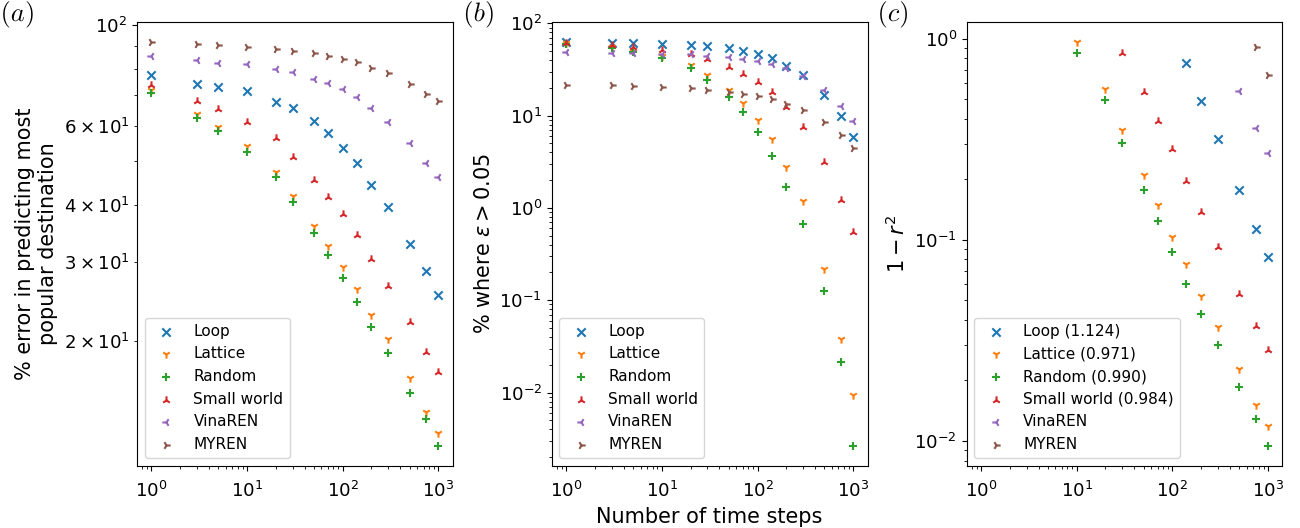}
\caption{a: The percentage error in predicting the most popular destination with respect to an origin, for various networks. Each plot point is the result of supplying as input data the last $T$ time steps of the number of agents on every ADE of the network. b: The corresponding plot showing the percentage of predictions where $\varepsilon>0.05$. c: The corresponding plot showing $1-r^2$, with $r^2$ being the coefficient of determination of fitting $\zeta_{ij, predicted}=\zeta_{ij, true}$. Shown in brackets in the legends of (c) are the exponents for the respective best fitted power laws. See Fig. 2 in SM for corresponding plots with lag of $l=10$ to traverse an edge.}
\label{fig6}
\end{figure*}
\begin{figure*}
\centering
\includegraphics[width=17cm]{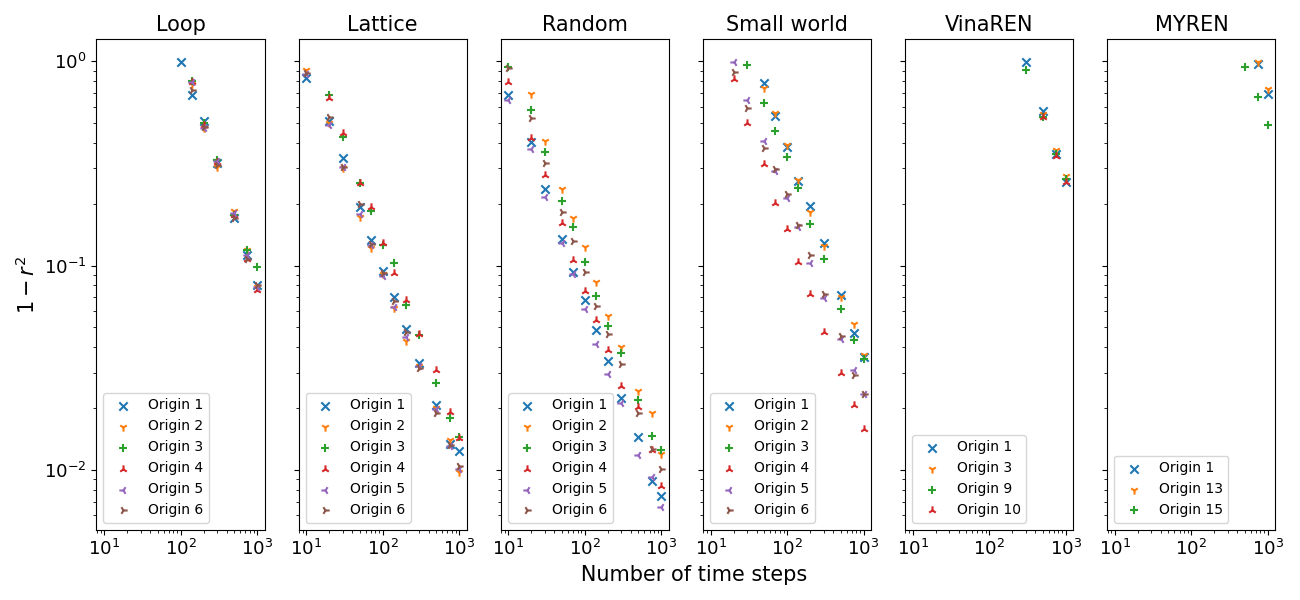}
\caption{In contrast to Fig.\ \ref{fig6} where each plot point is the average result over all origins, here are the corresponding results with respect to each origin, for $1-r^2$. For VinaREN and MYREN, many origin nodes are equivalent. Hence, there are only 4 distinct types of origins for VinaREN and 3 distinct types of origins for MYREN. See Figs.\ 3, 4, 5 in SM for the plots of $E$ and $\varepsilon$, as well as with lag $l=10$.}
\label{fig7}
\end{figure*}

\begin{figure*}
\centering
\includegraphics[width=13cm]{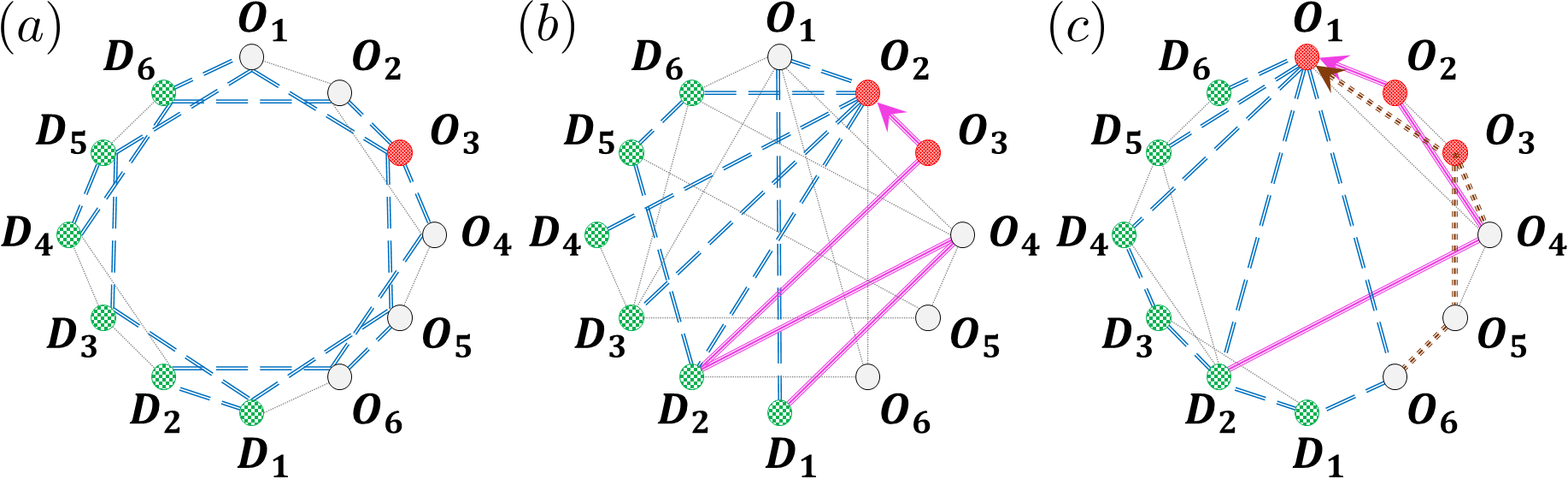}
\caption{The lattice, random and small-world networks from Figs.\ \ref{fig4}(b-d). Here, we highlight the origin nodes that are relatively worst in predicting the most popular destination, compared to other origin nodes. a: (lattice) $O_3$ and $O_4$ are the least accurate. Shown here is with respect to $O_3$, which is the furthest from all destination nodes. So, $O_3$ has many edges which overlap with the edges used by others. b: (random) Shown here are the edges used by $O_2$ to get to all destination nodes, with $O_3$ being a \emph{parasitic} origin node. This is because whilst $O_3$ uses $O_3\rightarrow D_2$ and $O_3\rightarrow D_2\rightarrow O_4\rightarrow D_1$, it otherwise overwhelmingly goes to $O_2$ and traverses all the edges used by $O_2$ to get to the destination nodes. c: (small-world) Shown here are the edges used by $O_1$ to get to all destination nodes, with $O_2$ and $O_3$ both being parasitic nodes as they overwhelmingly go to $O_1$ and traverse all the edges used by $O_1$ to get to the destination nodes.}
\label{fig8}
\end{figure*}

Consider a network with $M_O$ origin nodes and $M_D$ destination nodes. At each time step, origin $i$ can generate any number from 1 to $P$ agents each assigned to go to destination $j$ with probability $\zeta_{ij}$. Then, the agents propagate to subsequent edges in their paths as the time step progresses. Agents take the shortest path via edges of the network. If there are many shortest paths, one is randomly chosen (out of a maximum of $n$ pre-defined shortest paths). We consider two versions, viz. when an agent traverses an edge, it spends one time step before leaving that edge to another edge or arriving at its destination (without lag); or it spends $l$ time steps on the edge before proceeding (with lag). The number of agents on each edge is recorded every time step. The system is allowed to evolve, and the last $T$ time steps are taken to train a DNNGRU. 

We consider various such complex networks in our study (Fig.\ \ref{fig4}). As presented in the previous subsection where we compared with Vardi's EM algorithm, Fig.\ \ref{fig4}(a) is a directed loop with $M_O=M_D=6$. Then, we study three networks: Fig.\ \ref{fig4}(b) lattice, Fig.\ \ref{fig4}(c) random, Fig.\ \ref{fig4}(d) small-world, with the same number of origins and destinations ($M_O=M_D=6$) as well as edges (24 edges or 48 directed edges). These three networks provide a basis for comparison across different network topologies. Subsequently, we adopt two real-world internet networks \cite{InternetZoo}: Fig.\ \ref{fig4}(e) VinaREN, and Fig.\ \ref{fig4}(f) MYREN, to investigate the effects of real network topologies. The internet VinaREN and MYREN in Figs.\ \ref{fig4}(e, f) have $M_O=10,M_D=9$ and $M_O=16,M_D=15$, respectively. Note that the directed loop in Fig.\ \ref{fig4}(a) is also a real-world network of interest, and is studied in greater detail in Section \ref{scenarios}. In our study, we have set $P=10$, $n=4$, and $l\in\{1,10\}$. Also, the number of agents originating from each origin is changed every 20 time steps to introduce stochasticity and variation, though $\zeta_{ij}$ is kept constant throughout each dataset. The adjacency matrices for the networks in Figs. \ref{fig4}(b-d) and their set of shortest paths used from $O_i$ to $D_j$ are summarised in SM. Those for the other networks are deducible in a straightforward manner from Fig.\ \ref{fig4}. On top of these complex networks, we find it instructive to study a small lattice with $M_O=M_D=3$ (Fig.\ \ref{fig4}(g)), where we carry out an analytic treatment and compare against DNNGRU and LR.

We refer to the number of \emph{active directed edges} (ADE) as those links that are actually being used, i.e. forming part of the shortest paths from origins to destinations. The number of ADE for the networks in Figs.\ \ref{fig4}(a-g) are, respectively: 11, 38, 32, 24, 24, 40, 18. More ADE implies more information being tracked. Let the number of agents on every ADE at every time step form the feature dataset to be fed into the input layer of a DNNGRU, lasting $T$ time steps (c.f. Fig.\ \ref{fig2}). Unlike the previous subsection where we tested on Vardi's network, information of $\vec{V}$ is not supplied to DNNGRU, i.e. we no longer consider DNNGRU-V in the rest of this paper.

\subsubsection{A small lattice}

In order to gain a deeper understanding on the performance of LR and DNNGRU, we employ a small lattice (Fig. \ref{fig4}(g)) that is amenable to analytical treatment. The analytical treatment essentially tracks the paths of agents through the various edges, to arrive at a set of equations which exactly solves for all the unknown variables (elaborated in SM). Specifically, as the agents propagate from an origin node of the small lattice, those who are destined for the same end node may traverse different paths, and the travelling pattern of the agents can be modelled analytically. Nonetheless, there is still stochasticity in the analytical treatment due to the stochastic nature of agents spawning at the origins. And with the evaluation of $\zeta_{ij}$ based on the ratio of two integer number (of agents), there can still be uncertainties in the prediction of $\zeta_{ij}$. Note that the analytical treatment is achievable here due to the small lattice having few nodes with a relatively high number of edges. On the other hand, LR as given by Eq.\ (\ref{generalmultivariate}) is only an approximation as it does not track the correct number of $y_j$. Thus, we observe the better performance of the analytical treatment over LR in Figs.\ \ref{fig3}(g, h, i). DNNGRU matches with the analytical results, but appears to be slightly inferior in the asymptotic limit with large $T$, perhaps due to the optimisation of DNNGRU not finding the absolute best minimum of the loss.

\subsubsection{Topological effects of the complex networks}

Let us concern ourselves with the three networks: lattice;  random; and small-world (Figs.\ \ref{fig4}(b-d)), which have the same number of origin and destination nodes, and also the same number of edges. Due to the different topologies of these networks, they have distinct ADE and thus diverse shortest paths that the agents could travel from the origin node to destination node. As DNNGRU learns a model that takes these different shortest paths of the agents into account while LR does not, DNNGRU outperforms LR in prediction accuracy as shown in Fig.\ \ref{fig5}. (Note that LR is an exact model for the loop in Fig.\ \ref{fig4}(a), and we see convergence between LR and DNNGRU for large $T$ in the top row of Fig.\ \ref{fig5}.)

Fig.\ \ref{fig6} shows results of DNNGRU applied to the networks in Figs.\ \ref{fig4}(a-f), so we can compare the performances across various networks on the same plots. We examine how topology influences the performance of the three networks with equal number of nodes and edges using DNNGRU, i.e. those in Figs. \ref{fig4}(b-d). It turns out that the small-world topology performs the worst, given the same number of nodes and edges, because many shortest paths prefer taking the common “highway edges”, the very property that makes it small-world. The ramification of this is its lesser ADE (at 24) which carries less information than the lattice or random networks with greater ADE. This observation is, however, inconsistent with the fact that the random network which has a lesser ADE of 32 outperforms the lattice network with a higher ADE of 38. The deeper reason is revealed by looking into the performance variation across different origin nodes within the network.

Recall that each origin node $i$ is trained with its own DNNGRU to exclusively predict $\zeta_{ij}$ for that $i$. Fig.\ \ref{fig7} reports the results for $1-r^2$ versus $T$ with respect to each particular origin node. Corresponding results for the percentage errors of predicting the most popular destination as well as those for $\varepsilon>0.05$ are given in SM.

Since the lattice, random and small-world networks have origin nodes with different topological properties, the agents could have alternative pre-defined shortest paths to take. Consequently, different network topologies do indeed lead to different origin nodes performing better than others. We can determine the relatively worst performing origin nodes in these networks, by examining the graphs in Fig.\ \ref{fig7}. For the lattice, origin nodes $O_3$ and $O_4$ (which are equivalent, due to the symmetry of the lattice) have a relatively higher error compared to the other origin nodes. For the random network, the worst performing origin nodes are $O_2$ and $O_3$ with $O_6$ also relatively poor, whilst those for the small-world network are $O_1$, $O_2$ and $O_3$.

Accuracy in inferring $\zeta_{ij}$ relies on the ability to disentangle the individual $i,j$ components from the combined information when they add up on the shared edges. The lattice in Fig.\ \ref{fig8}(a) shows how $O_3$ (and $O_4$, by symmetry) tends to be inferior compared to the rest, because it is the furthest away from all destination nodes. This means that it has to traverse relatively more edges, and consequently overlap with more edges used by other origin nodes.

Another way an origin node can become inferior is due to a \emph{parasitic origin node} that taps on it to get to the destinations. This is obvious in the random and small-world networks, as depicted in Figs.\ \ref{fig8}(b, c). For the random network, $O_3$ has essentially all its paths relying on all those used by $O_2$. This makes both of them suffer slightly worse performance, since it is more difficult to ascertain the $\zeta_{ij}$ to be attributed to which of them. Other origin nodes have more diverse paths, allowing for more information available to infer their own $\zeta_{ij}$. Similarly for the small-world network, $O_2$ and $O_3$ are parasitic origin nodes with majority of their paths going via those of $O_1$, resulting in all three of them being slightly inferior than other origin nodes.

Let $E_{ij}$ be the total number of edges that origin nodes $O_i$ and $O_j$ would overlap, when getting to all destinations. We summarise the values of $E_{ij}$ for these three networks in Table\ \ref{table1}. Generally, origin nodes that have greater overlap, i.e. larger $E_{ij}$, would tend to be less accurate. These can be made more precise by encapsulating them into an index $\chi_i$ for each origin node $O_i$ which quantifies the relative degree of overlap amongst them:
\begin{align}\label{chi}
\chi_i&=\frac{1}{E_i(M_O-1)}\sum_{i\neq j}{E_{ij}^2}.
\end{align}
In Eq.\ (\ref{chi}), $E_i$ is the number of edges that are involved in getting from $O_i$ to all destinations. The division by $M_O-1$ serves to make $\chi_i$ as an averaged quantity over all the overlapping origin nodes (minus one to exclude itself). The division by $E_i$ is to normalise as nodes with more edges would tend to proportionally overlap more with other origins' edges. The rationale for taking the sum of squares of $E_{ij}$ (as opposed to just the sum of $E_{ij}$) is because an edge overlap involves two such origin nodes $O_i$ and $O_j$, hence $E_{ij}$ should arguably appear as two factors. The squaring would also more greatly penalise higher overlapping numbers as compared to fewer overlaps. This is especially critical for an origin node with its parasitic companion which together share a larger amount of edges $E_{ij}$, and which would raise both their indices values $\chi_i$ and $\chi_j$ by $\sim E_{ij}^2$.

\begin{table*}
\centering
\begin{tabular}{ccc}
Lattice & Random & Small-world\\
\begin{tabular}{|c|c|c|c|c|c|c|c|}
\hline
Node & $O_1$ & $O_2$ & $O_3$ &$O_4$ & $O_5$ & $O_6$ & $\chi_i$\\
\hline
$O_1 (11)$ & --- & 6 & 7 & 3 & 3 & 0 & 1.87\\
$O_2 (16)$ & 6 & --- & 8 & 10 & 4 & 3 & 2.81\\
\color{red}{$O_3 (16)$} & \color{red}{7} & \color{red}{8} & \color{red}{---} & \color{red}{8} & \color{red}{10} & \color{red}{3} & \color{red}{3.58}\\
\color{red}{$O_4 (16)$} & \color{red}{3} & \color{red}{10} & \color{red}{8} & \color{red}{---} & \color{red}{8} & \color{red}{7} & \color{red}{3.58}\\
$O_5 (16)$ & 3 & 4 & 10 & 8 & --- & 6 & 2.81\\
$O_6 (11)$ & 0 & 3 & 3 & 7 & 6 & --- & 1.87\\
\hline
\end{tabular}&
\begin{tabular}{|c|c|c|c|c|c|c|c|}
\hline
Node & $O_1$ & $O_2$ & $O_3$ &$O_4$ & $O_5$ & $O_6$ & $\chi_i$\\
\hline
$O_1 (12) $ & --- & 4 & 2 & 5 & 2 & 5 & 1.23\\
\color{red}{$O_2 (8) $} & \color{red}{4} & \color{red}{---} & \color{red}{6} & \color{red}{3} & \color{red}{0} & \color{red}{5} & \color{red}{2.15}\\
\color{red}{$O_3 (10) $} & \color{red}{2} & \color{red}{6} & \color{red}{---} & \color{red}{3} & \color{red}{1} & \color{red}{5} & \color{red}{1.50}\\
$O_4 (14) $ & 5 & 3 & 3 & --- & 6 & 3 & 1.26\\
$O_5 (9) $ & 2 & 0& 1 & 6 & --- & 0 & 0.91\\
\color{red}{$O_6 (10) $} & \color{red}{5} & \color{red}{5} & \color{red}{5} & \color{red}{3} & \color{red}{0} & \color{red}{---} & \color{red}{1.68}\\
\hline
\end{tabular}&
\begin{tabular}{|c|c|c|c|c|c|c|c|}
\hline
Node & $O_1$ & $O_2$ & $O_3$ &$O_4$ & $O_5$ & $O_6$ & $\chi_i$\\
\hline
\color{red}{$O_1 (9) $} & \color{red}{---} & \color{red}{9} & \color{red}{9} & \color{red}{5} & \color{red}{5} & \color{red}{5} & \color{red}{5.27}\\
\color{red}{$O_2 (12) $} & \color{red}{9} & \color{red}{---} & \color{red}{10} & \color{red}{6} & \color{red}{6} & \color{red}{5} & \color{red}{4.63}\\
\color{red}{$O_3 (14) $} & \color{red}{9} & \color{red}{10} & \color{red}{---} & \color{red}{6} & \color{red}{8} & \color{red}{5} & \color{red}{4.37}\\
$O_4 (9) $ & 5 & 6 & 6 & --- & 8 & 3 & 3.78\\
$O_5 (15) $ & 5 & 6 & 8 & 8 & --- & 5 & 2.85\\
$O_6 (8) $ & 5 & 5 & 5 & 3 & 5 & --- & 2.73\\
\hline
\end{tabular}
\end{tabular}
\caption{Crosstables for the lattice, random and small-world networks from Figs.\ \ref{fig4}(b-d). The numbers in parentheses in the left most column of each crosstable indicate the total number of edges $E_i$ involved for that node to get to all destinations. These crosstables show the number of edges $E_{ij}$ that origin node $O_i$ overlaps with those of origin node $O_j$, when getting to any destination. Rows coloured in red show the poorer or poorest performing nodes for that network, as deduced from Fig.\ \ref{fig7}. As revealed by the crosstables, these correspond to large values of $\chi_i$, signifying greater overlaps.}\label{table1}
\end{table*}

As revealed in Table\ \ref{table1}, larger values of $\chi_i$ would correspond to that origin node $O_i$ being less accurate (or larger error) relative to other origin nodes within that network, as deduced from Fig.\ \ref{fig7}. Incidentally, this index $\chi_i$ also appears to correspond to the random network performing better than the lattice, with the small-world network as the worst. The average values $\bar{\chi}$ of $\chi_i$ for these networks (random, lattice, small-world) are $1.46,2.75,3.94$, respectively. This was, in fact, already deduced from Fig.\ \ref{fig6} earlier. Intriguingly, the average index $\bar{\chi}$ explains why the random network with 32 ADE manages to outperform the lattice with 38 ADE: The former's average $\bar{\chi}$ is smaller which indicates less overlapping of edges used between the origin nodes. Thus, by accounting for the overlapping of edges as a measure of how hard it is to disentangle which agents are from which origins to which destinations, we obtain an indicator of which origin nodes are easier to predict the OD probabilities. On the other hand, we are able to verify this empirically using the model-free DNNGRU on each origin node (as displayed in Fig.\ \ref{fig7}).

\subsubsection{No overlap, partial overlap, superhighway networks}

\begin{figure*}
\centering
\includegraphics[width=17cm]{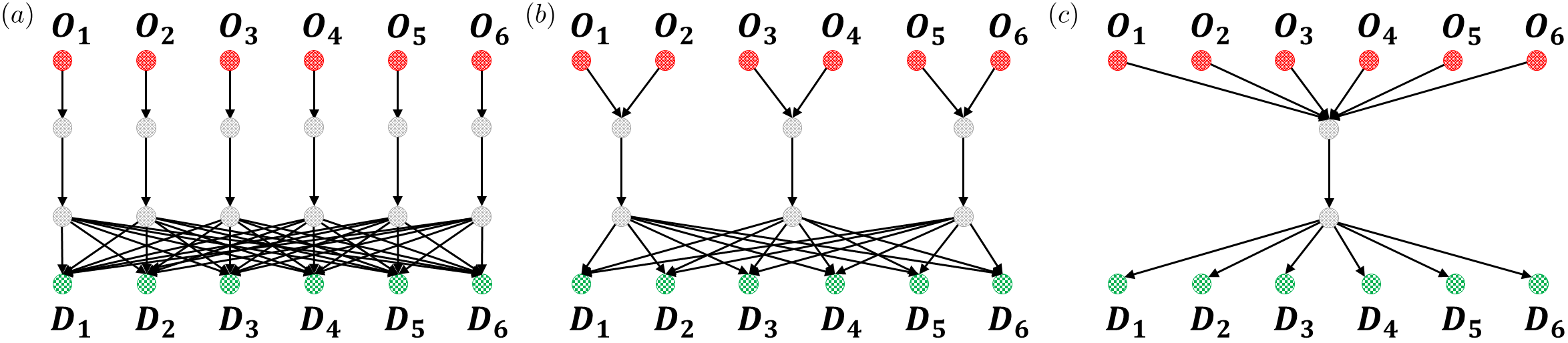}
\caption{Three networks with 6 origins and 6 destinations, with differing levels of overlap. a: No overlap. b: Partial overlap. c: Superhighway.}
\label{fig9}
\end{figure*}

In this subsection, we provide a direct illustration of the effects of topology and sharing of edges on the accuracy of predicting OD information through a study of idealised networks as shown in Fig.\ \ref{fig9}. Here, three network with six origins and six destinations are considered. The first has no overlap, so inferring $\zeta_{ij}$ from the edge counts does not contain the uncertainties of the origin or destination on which the agent traverses, though there is still stochasticity involved since agents choose their destinations probabilistically. The second network has partial overlap, since two origin nodes share a common edge. The inference of $\zeta_{ij}$ can be achieved by simple linear regression involving the pair of nodes, based on Eq.\ (\ref{generalmultivariate}). Finally, the third network has all six origin nodes sharing a common superhighway edge. Similar to the second network, here linear regression is directly applicable with all six origin nodes.

\begin{figure*}
\centering
\includegraphics[width=17cm]{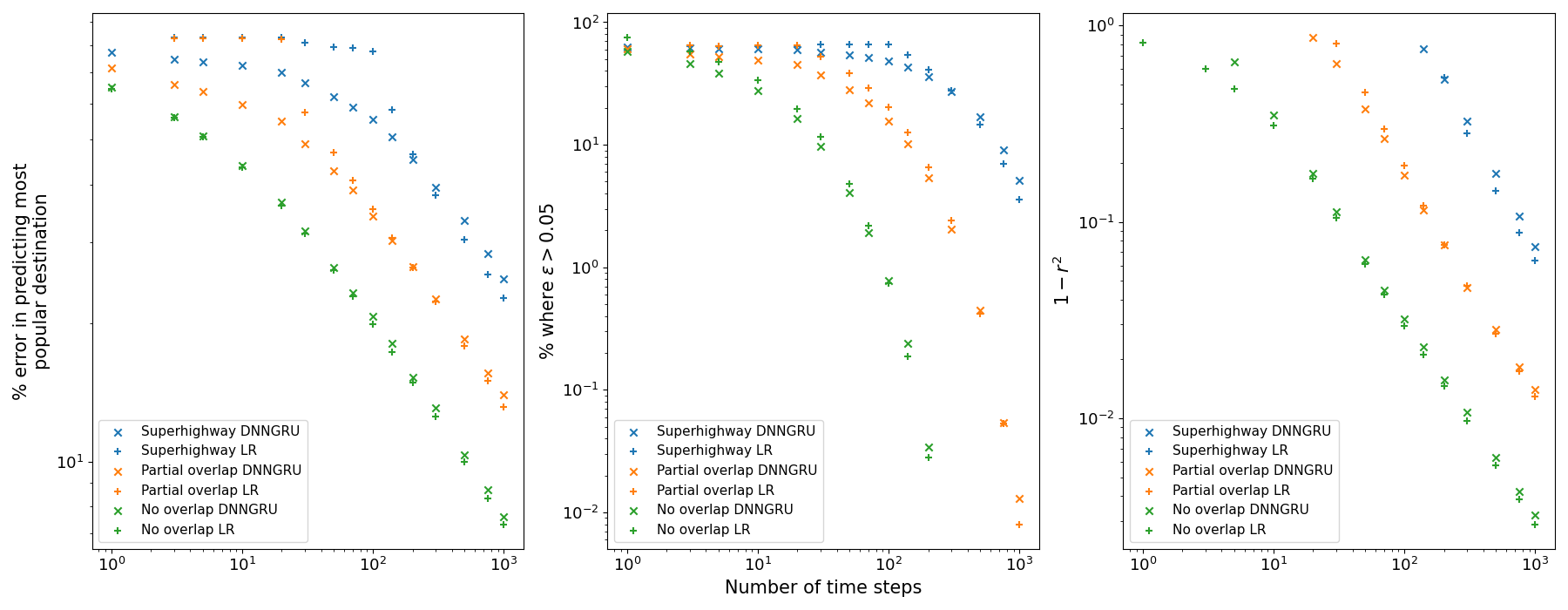}
\caption{Performance comparison between DNNGRU and LR for the three networks in Fig.\ \ref{fig9}.}
\label{fig10}
\end{figure*}

The results are shown in Fig.\ \ref{fig10} . It is evident that no overlap is the easiest to infer $\zeta_{ij}$, with the superhighway network being the hardest. Because LR provides an exact formulation of the three networks, it consistently performs the best given sufficient data. In all the three networks, we observe that accuracy improves in a power law manner with more data, with DNNGRU matching the results of LR. We can calculate $\bar{\chi}$ for each of these networks. The index value for the no overlap network is zero, since $E_{ij}=0$ for all pairs of origins. For the partial overlap network, the sum of $E_{ij}^2$ in Eq.\ (\ref{chi}) is only $7^2$ as each origin only overlaps with one other origin. Finally, the sum of $E_{ij}^2$ for the superhighway network is $5\times7^2$ since each origin overlaps with five other origins. Consequently, $\bar{\chi}=0,1.225,6.125$, respectively for the no overlap, partial overlap and superhighway networks. This again shows $\bar{\chi}$ as a measure of the degree of shared edges, with higher values signifying lower accuracy. Nevertheless, note from Fig.\ \ref{fig10} that increasing the data size would improve the prediction accuracy in a power law manner in all cases, even for a superhighway network where all information pass through a common edge.



\subsubsection{Real-world complex networks}

Let us now consider the topology of three real-world complex networks: the directed loop, VinaREN, and MYREN. The directed loop corresponds to a bus transport network that provides loop service with $M_O=M_D=6$ origin and destination bus stops. In comparison to the topology of the networks of the last section which have the same number of origin and destination nodes, the directed loop with only 11 ADE performs worst. This results from a larger average overlapped index $\bar{\chi}=6.51$ due to a greater number of overlapped edges and parasitic origin nodes which implies poorer accuracy as it is harder to disentangle the various ODs. Nonetheless, the directed loop is the only network amongst those in Figs.\ \ref{fig4}(a-d) where all origin nodes display essentially equivalent performance. The indices $\chi_i$ for these origins are $6.00,6.60,6.91,6.93,6.63,6.00$, respectively, which are highly similar and consistent with them having comparable accuracies. The directed loop offers no alternative shortest path, so every agent from origin $i$ to destination $j$ takes the same path. The application of the directed loop for a bus loop service will be detailed in the next section.

For the internet networks VinaREN and MYREN, they have more destinations $M_D$ to deal with, and generally report larger errors in predicting the most popular destination. Their ADE are comparable to the smaller networks in Figs.\ \ref{fig4}(b-d), such that they do not quite possess proportionately sufficient ADE to deal with more nodes. This is a consequence of the features that many real-world internet networks are scale-free and small-world \cite{Watts98,Latora01}. Whilst these two networks are relatively small to be considered anywhere near being scale-free, this is nonetheless plausibly the situation with scale-free networks which are ultra small \cite{Chung02,Cohen03}: Scale-free networks probably have comparably poorer performance due to the ultra few ADE. In the case of VinaREN and MYREN, this small-world feature leads to the presence of common highway edges which cause considerable overlapping of edges. This explains the worst performance of VinaREN and MYREN compared to the rest of the network topologies. In addition, because of the presence of alternative shortest paths in the MYREN network relative to the VinaREN network, which introduces additional uncertainties, the MYREN network performs worse compared to VinaREN.

\subsection{Application: Bus loop service}\label{scenarios}

\begin{figure*}
\centering
\includegraphics[width=17cm]{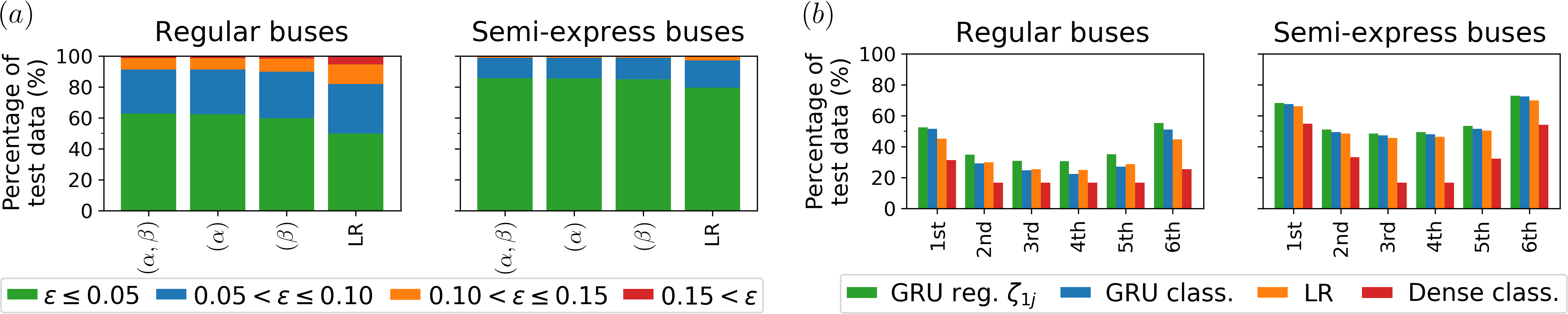}
\caption{a: Regression: Predicting $\zeta_{1j}$ by DNNGRU using ($\alpha,\beta$), $\alpha$, $\beta$; and by LR using $\alpha$. b: Classification: Predicting the most, second most, $\cdots$, least likely destinations from origin $i=1$ by three different DNN architectures, and LR using $\alpha$.}
\label{fig11}
\end{figure*}

We consider a loop of $M_O$ origin bus stops with $M_D$ destination bus stops served by $N$ buses in the form of Fig. \ref{fig4}(a). Without loss of generality, origins are placed before destinations, all staggered. Regular buses go around the loop serving all bus stops sequentially. To describe other complex network topologies in a bus system, we additionally implement a \emph{semi-express} configuration where buses only board commuters from a subset of origins whilst always allow alighting \cite{Aramsiv21}. This configuration turns out to be chaotic, but outperforms regular or fully express buses in minimising commuters waiting time with optimal performance achieved at the edge of chaos \cite{Vee2021}. Let $N=M_O=M_D=6$. This system represents a morning commute scenario where students living in the NTU campus residences commute to faculty buildings (Fig.\ \ref{fig1}(d)). (The map shows $7$ residences and $5$ faculty buildings. We approximate this with $M_O=M_D=6$.) The complex network for semi-express buses (carriers, i.e. edges) are depicted in Fig.\ \ref{fig1}(e). Each semi-express bus boards people from three origins, then allows alighting at every destination. Different semi-express buses do not board from the same origin subset.

Unlike edges in general complex networks, each bus in a bus network maintains its identity as a carrier between nodes. In other words, passengers from different buses do not mix even if the buses share the same road. On the other hand for general complex networks, an edge connecting a pair of nodes would combine all agents that traverse it. This makes bus systems relatively easier to be analytically treated, and so we can compare performances by DNNGRU and LR.

We can simulate each of these two bus systems given some $\zeta_{ij}$ to generate: $\alpha$) the number of commuters on the buses after leaving a bus stop; and $\beta$) the duration that the buses spend at bus stops. Our parameters are based on real data measured from NTU campus shuttle buses \cite{Vee2019,Vee2019b,busurl}. We record 10 laps of service for each bus. This is reasonable as a bus takes $\sim20$ minutes a loop corresponding to $3.5$ hours of service for the morning commute from $8$ am to $11\textrm{:}30$ am. All buses' time series are concatenated into one single feature, for each $\alpha$ and $\beta$. Empirically, this longer time series turns out to significantly improve performance, compared to each bus representing separate features. As before, DNN training comprises 250k datasets, with another 10k as validation during training. Additionally, 50k datasets are for testing. The latter are also used by LR to predict $\zeta_{ij}$. The various DNN architectural details are given in Methods.

With respect to $i=1$  (other origins $i\neq1$ to be trained separately), DNNGRU predicts $\zeta_{1j}$ using ($\alpha,\beta$), $\alpha$, or $\beta$. LR predicts $\zeta_{ij}$ for all $i$, using $\alpha$ where $(x_i,y_j)$ are directly deducible, and from which we take $\zeta_{1j}$ for comparison. Fig.\ \ref{fig11}(a) shows the performance of DNNGRU clearly outperforming LR in predicting $\zeta_{1j}$. The green, blue, orange, red bars respectively represent $\varepsilon\leq0.05$, $0.05<\varepsilon\leq0.10$, $0.10<\varepsilon\leq0.15$, $0.15<\varepsilon$, where $\varepsilon$ is the difference between the predicted $\zeta_{1j}$ and the true $\zeta_{1j}$. Generally, using $\alpha$, $\beta$ or both for DNNGRU are equally good. A semi-express network topology has higher prediction accuracy than that of regular buses since it allows for more diverse data to learn from. The former topology is also a more efficient bus system \cite{Aramsiv21}, and conceivably in general complex networks as it provides greater variety of pathways than one single loop (as discussed in Subsection \ref{GCN}).

From $\zeta_{1j}$ obtained by DNNGRU and LR using $\alpha$, we order the destination ranking for origin $i=1$ according to their probabilities and display their prediction accuracies in Fig.\ \ref{fig11}(b). These are compared with directly classifying destination ranking by GRU or dense layers using $\alpha$. Direct classification requires training individual DNNs for each rank, i.e. one DNN learns to predict the most likely destination, another learns to predict the second most likely one, etc. Despite having dedicated DNNGRUs to classify each rank, they turn out to always be inferior to one DNNGRU predicting $\zeta_{1j}$ to then obtain the ranking. Intriguingly, they are only marginally superior (but not always) to LR. Consequently, if there are insufficient training data for DNNGRU, LR serves as a quick estimate with respectable accuracy compared to DNNGRU. Dense DNNs are consistently poorest due to its simplicity in not inferring temporal information in the time series, and also nowhere comparable to LR. DNNGRU regressing $\zeta_{1j}$ is always the best. Architectural details on using DNNGRU for direct classification and dense DNN for classification are given in Methods.

Classifying destination ranking produces a $U$-shape, with predicting the least likely destination being most accurate. This is generally true for all the networks studied in Fig.\ \ref{fig4} as well, when trying to predict the other rankings other than the most popular destination. Errors in predicting $\zeta_{1j}$ may mess up the ordering, and those in between ($2$nd most likely, $\cdots$, $2$nd least likely) can be affected both above and below. The most and least likely ones are only affected from below or above, respectively, with the latter being bounded by zero --- giving it slightly better accuracy. Incidentally, DNN direct classification generally tends to overfit from excessive training, with test accuracy systematically less than train accuracy. Conversely, DNN regressing for $\zeta_{1j}$ does not seem to suffer from overfitting with test accuracy remaining comparable to train accuracy.

These results are useful for practical applications, as it informs us which bus stops should be prioritised and which may be skipped if needed. Train loops are another common loop services. For instance, major cities in the world have light rail system connected to rapid transit system. Unlike buses, trains have dedicated tracks free from traffic. Therefore, they are cleanly scheduled to stop over prescribed durations and do not experience bunching. This implies $\beta$ is not applicable, as the duration a train spends at a station is not proportional to demand. Nevertheless, $\alpha$ is measurable for inferring the OD distribution $\zeta_{ij}$ of this train loop. Moreover, we can use DNN with supervised learning which is network-free on routes with multiple loops and linear/branching topologies.

\section{Discussion}\label{discussion}

Knowledge of $\zeta_{ij}$ is akin to possessing knowledge of the dynamical law of the system. With respect to this dynamical law, we can figure out optimal configurations of buses for delivery of commuters from their origins to desired destinations. As a concrete implementation, the latter is achieved via multi-agent reinforcement learning (MARL) recently carried out \cite{Aramsiv21}. However, arbitrary $\zeta_{ij}$ were tested there like uniform distribution (fixed proportion of commuters alighting at every destination) or antipodal (commuters alight at the destination opposite where they boarded on the loop). More complicated $\zeta_{ij}$ can certainly be prescribed for the MARL framework in Refs.\ \cite{Vee2019d,Aramsiv21}, but the most useful one would be the actual $\zeta_{ij}$ corresponding to the bus loop service being studied, like our NTU campus loop shuttle bus service \cite{Vee2019,Vee2019b,Quek2020}. Thus, the algorithm presented here is complementary to the MARL framework in Ref.\ \cite{Aramsiv21}. We intend to further implement our algorithm to city-wide bus networks from publicly accessible data \cite{busurl} with MARL optimisation, to be reported elsewhere. Other complex systems (c.f. Fig.\ \ref{fig1}) can be similarly modelled towards improving the efficiency of agent-transfer or impeding undesirable transactions.

In studying general complex networks, we revealed how different network topologies can lead to (dis)advantageous performances, quantifying in terms of $\chi_i$ and the presence of alternative shortest paths. We also studied how different origin nodes' properties may make them perform slightly better than other origin nodes, in the examples of the lattice, random, small-world networks, as well as real-world networks like the loop, VinaREN and MYREN. Notably, we demonstrated empirically that the use of recurrent neural network architecture like GRU allows for longer time-series data to generally yield more accurate predictions, with the time-series data length and prediction error appearing to scale as a power law.

Whilst we have illustrated a concrete application in a university bus loop service, the framework presented here can be used in various other areas like Internet tomography \cite{VZ80,Vardi96,Tebaldi98,Cao00,Coates02}, city-wide bus/traffic networks \cite{Latora01,Latora02,Barry02,Barry09,Zhao07,Trepanier07,Farzin08,Nassir11,Gordon13,Muni12,Nunes16,Hora17,Tebaldi98,Hazelton00,Bera11,Cascetta13,Yang17,Bauer18,Li18,Dragu19,Dey20}, as well as in mapping global epidemic/pandemic propagation and contact tracing \cite{Bonabeau02,Balcan09,Zhao15,Weso15,Weso16,Gomez19,Grantz20,Buckee20,Pepe20,Cintia21,Ciavarella21,Murphy21}. As we have shown how DNNGRU and our formulation of linear regression consistently outperform existing methods using expectation-maximisation with moments \cite{Vardi96}, Bayesian inference \cite{Tebaldi98} and entropy maximisation \cite{VZ80}, we expect further impactful advancements in OD inference based on this work. With DNNGRU being network-free where it only requires training it with data by supervised learning with no requirement of analytical modelling, this approach is scalable. We envision significant improvements in OD mapping in these other fields, bringing with them major enhancements whilst respecting privacy.

\section{Methods}

\subsection{DNNGRU hyperparameters}

We employ a deep neural network (DNN) with two hidden layers consisting of gated recurrent units (GRU). The input layer comprises the measurement of the number of agents on each active directed edge (ADE) of a complex network at some moment in time. If there are $L$ ADE, then there are $L$ units in the input layer. As the next two (hidden) layers are GRUs stacked on one another, the input data to the input layer can contain $T$ time steps. In other words, each of the $L$ units in the input layer comprises a time series of length $T$. Finally, the output layer comprises $M_D$ units (recall that $M_D$ is the number of destination nodes), each giving the probability of agents from some specific origin node $i$ to end up at destination $j$. Different DNNs are trained individually for different origin nodes $i$. (See Fig.\ \ref{fig2}.)

For a given setup, the input layer and output layer have fixed number of units. So, the number of GRU hidden layers and the number of units in each GRU hidden layer are user-defined. We fix each GRU hidden layer to have 128 units. We observe that increasing the width of the hidden layers generally improve performance. However, the number of trainable weights and biases in the DNNGRU would rapidly blow up with increasing width size. The choice of 128 units for each hidden GRU layer balances this, with $\sim150$k trainable weights and biases, which can be well-trained by 250k training datasets. On the other hand, whilst two hidden layers definitely outperform one hidden layer, three or more hidden layers do not show improved performance over two hidden layers. Thus, two hidden layers seem optimal.

DNNGRU training is carried out on TensorFlow 2.3 using Keras \cite{keras}. Default activation function is used for GRU, whilst the output layer uses sigmoid. Although the sum of all output values is $1$ (c.f. the summation equation below Eq.\ (\ref{generalmultivariate})), softmax does not seem desirable as it sometimes leads to stagnant training. The loss function used is binary cross entropy, optimised with the standard ADAM. An $L^2$ recurrent regularisation is applied on the GRU hidden layers. This prevents blowing up of the weights which occasionally occurs during training if no regularisation is used. Batch size of 128 is used, which is generally better than 64 or 32. However, larger batch sizes would tax the GPU VRAM, leading to sporadic GPU breakdowns. Training is carried out over 200 epochs. Generally, optimal performance is achieved well before 200 epochs and continues to incrementally improve. No significant overfitting is observed, so no early stopping is implemented. In fact, for many of the DNNs, training accuracy is essentially similar to validation accuracy. Sometimes, the latter is slightly less than the former, but typically continues to improve over training epochs. Incidentally, input quantities which are small ($\ll100$) are left as they are, whilst larger input values ($\gtrsim100$) are better to be rescaled by appropriate division, so that the DNN performs optimally.

\subsection{Transfer learning}

With GRU layers, the input layer can take any time series length $T$ and as long as the number of ADE $L$ remains the same, then the DNNGRU architecture remains the same. A longer $T$ would require longer training time since the time series data are processed sequentially by the DNNGRU and cannot be parallelised. This unmodified DNNGRU architecture regardless of $T$ allows the implementation of transfer learning when supplying inputs of different temporal lengths $T$ to obtain each plot point in Figs.\ \ref{fig3}, \ref{fig5}, \ref{fig6}, \ref{fig7}. So with respect to each origin node $i$, a DNNGRU is trained completely from scratch with $T=1000$. After this has completed, the trained weights are used as initial weights for the next shorter time series with $T=750$, and trained for only 20 epochs instead of the full 200 epochs. This generally leads to optimised performance similar to that from complete training with random initial weights (we carried out some tests and this is generally true), but only takes $10\%$ of training time. Then, the trained weights for $T=750$ are used as initial weights to train a DNNGRU for $T=500$, and so on until $T=1$.

\subsection{The coefficient of determination \texorpdfstring{$r^2$}{r2}}

The coefficient of determination, $r^2$ is given by
\begin{align}\label{COD}
r^2&=1-\frac{\sum{(\zeta_{ij, predicted}-\zeta_{ij, true})^2}}{\sum{(\zeta_{ij, predicted}-\overline{\zeta_{ij, predicted}})^2}},
\end{align}
where $\overline{\zeta_{ij, predicted}}$ is the mean of all $\zeta_{ij, predicted}$. This imposes an exact fit of $\zeta_{ij, predicted}=\zeta_{ij, true}$ with zero intercept, as opposed to the correlation coefficient of a linear regression fit that allows for an unspecified intercept to be determined from the fitting. Note that the coefficient of determination $r^2$ can take negative values, which would imply that the predictions are worse than the baseline model that always predicts the mean $\overline{\zeta_{ij, predicted}}$ (i.e. $\zeta_{ij, true}=\overline{\zeta_{ij, predicted}}$, giving $r^2=0$). Hence, $r^2<0$ are not shown in the plots in Figs.\ \ref{fig3}, \ref{fig5}, \ref{fig6}, \ref{fig7}, as they imply that the predictions are just so poor, they are essentially nonsensical. In contrast, $r^2=1$ implies perfect predictions (since \emph{every} $\zeta_{ij, predicted}=\zeta_{ij, true}$).

\subsection{Vardi's expectation-maximisation algorithm}

We present an analytical comparison between linear regression (LR) and Vardi's expectation-maximisation (EM), to elucidate how a switch in paradigm from trying to infer actual whole numbers of the OD matrix to instead inferring the probabilities would lead to significant improvements in the predictions. It is instructive to consider a directed loop for this purpose as shown in Fig.\ \ref{fig4}(a), which is also what happens when a bus picks up people from $M_O$ origin bus stops and then delivers them at $M_D$ destination bus stops (c.f. Fig.\ \ref{fig1}(d)). This bus system is the subject of Section \ref{scenarios}, which is based on the NTU shuttle bus service \cite{Vee2019}.

The bus would pick up $x_i$ commuters from each of the origin bus stops ($i=1,\cdots,M_O$), and then let $y_j$ commuters alight at each of the destination bus stops ($j=1,\cdots,M_D$) according to $\zeta_{ij}$, which leads to Eq.\ (\ref{generalmultivariate}). In other words, everybody who alights at destination bus stop $j$ for each $j=1,\cdots,M_D$ comprises everybody who boarded from each of the origin bus stops who wants to go to bus stop $j$. Since we can repeatedly measure all the $x_i$ and $y_j$ as the bus loops around over time, Eq.\ (\ref{generalmultivariate}) is naturally in a form where LR is directly applicable. The OD probabilities $\zeta_{ij}$ are coefficients of a multivariate system of linear equations, and the best set of values of $\zeta_{ij}$ is the one that would minimise the mean squared error of the predicted $y_j$ given the $x_i$ according to Eq.\ (\ref{generalmultivariate}), versus the measured $y_j$.

On the other hand, Vardi \cite{Vardi96} approaches the problem differently. Instead of evaluating the OD number of commuters from the OD probabilities $\zeta_{ij}$, Vardi determines the actual OD whole numbers of commuters by modelling them directly as Poisson random variables with mean $\lambda_k$, where $k=1,\cdots,M_OM_D$. With the correct Poisson parameters $\lambda_k$ representing the mean (and variance) of each OD from some $O_i$ to some $D_j$, the actual number of people who want to go from $O_i$ to $D_j$ is the random variable $X_k\sim\textrm{Poisson}(\lambda_k)$.

Without loss of generality, let us fix $M_O=M_D=6$ as in Fig.\ \ref{fig4}(a). This gives the following matrix equation:
\begin{align}\label{YAX}
\vec{Y}&=A\vec{X}.
\end{align}
The column vector $\vec{X}$ is the collection of all 36 possible OD for this network. More specifically, we define
\begin{widetext}
\begin{align}
\vec{X}=( X_{O_1D_1},X_{O_1D_2},X_{O_1D_3},X_{O_1D_4},X_{O_1D_5},X_{O_1D_6},X_{O_2D_1},X_{O_2D_2},\cdots,X_{O_6D_6})^T,
\end{align}
\end{widetext}
with $X_{O_iD_j}$ denoting the actual number of people who go from $O_i$ to $D_j$. The column vector $\vec{Y}$ comprises the number of commuters on each of the ADE, i.e. the 11 edges that are being used in this loop: $O_1O_2,O_2O_3,\cdots,D_5D_6$. Explicitly,
\begin{align}
\vec{Y}=(Y_{O_1O_2}, Y_{O_2O_3},\cdots,Y_{D_5D_6})^T,
\end{align}
where $Y_{O_1O_2}$ is the number of people on the directed edge $O_1O_2$, etc. Then, $A$ is an $11\times36$ matrix that relates the vector $\vec{X}$ of number of people for each OD to the vector $\vec{Y}$ of number of people on each ADE. This matrix $A$ is referred to as the ``routing matrix''. For this loop network, we can determine $A$ to be:
\begin{widetext}
\begin{eqnarray}
A=
\begin{pmatrix}
1 & 1 & 1 & 1 & 1 & 1 & 0 & 0 & 0 & 0 & 0 & 0 & 0 & 0 & 0 & 0 & 0 & 0 & 0 & 0 & 0 & 0 & 0 & 0 & 0 & 0 & 0 & 0 & 0 & 0 & 0 & 0 & 0 & 0 & 0 & 0\\
1 & 1 & 1 & 1 & 1 & 1 & 1 & 1 & 1 & 1 & 1 & 1 & 0 & 0 & 0 & 0 & 0 & 0 & 0 & 0 & 0 & 0 & 0 & 0 & 0 & 0 & 0 & 0 & 0 & 0 & 0 & 0 & 0 & 0 & 0 & 0\\
1 & 1 & 1 & 1 & 1 & 1 & 1 & 1 & 1 & 1 & 1 & 1 & 1 & 1 & 1 & 1 & 1 & 1 & 0 & 0 & 0 & 0 & 0 & 0 & 0 & 0 & 0 & 0 & 0 & 0 & 0 & 0 & 0 & 0 & 0 & 0\\
1 & 1 & 1 & 1 & 1 & 1 & 1 & 1 & 1 & 1 & 1 & 1 & 1 & 1 & 1 & 1 & 1 & 1 & 1 & 1 & 1 & 1 & 1 & 1 & 0 & 0 & 0 & 0 & 0 & 0 & 0 & 0 & 0 & 0 & 0 & 0\\
1 & 1 & 1 & 1 & 1 & 1 & 1 & 1 & 1 & 1 & 1 & 1 & 1 & 1 & 1 & 1 & 1 & 1 & 1 & 1 & 1 & 1 & 1 & 1 & 1 & 1 & 1 & 1 & 1 & 1 & 0 & 0 & 0 & 0 & 0 & 0\\
1 & 1 & 1 & 1 & 1 & 1 & 1 & 1 & 1 & 1 & 1 & 1 & 1 & 1 & 1 & 1 & 1 & 1 & 1 & 1 & 1 & 1 & 1 & 1 & 1 & 1 & 1 & 1 & 1 & 1 & 1 & 1 & 1 & 1 & 1 & 1\\
0 & 1 & 1 & 1 & 1 & 1 & 0 & 1 & 1 & 1 & 1 & 1 & 0 & 1 & 1 & 1 & 1 & 1 & 0 & 1 & 1 & 1 & 1 & 1 & 0 & 1 & 1 & 1 & 1 & 1 & 0 & 1 & 1 & 1 & 1 & 1\\
0 & 0 & 1 & 1 & 1 & 1 & 0 & 0 & 1 & 1 & 1 & 1 & 0 & 0 & 1 & 1 & 1 & 1 & 0 & 0 & 1 & 1 & 1 & 1 & 0 & 0 & 1 & 1 & 1 & 1 & 0 & 0 & 1 & 1 & 1 & 1\\
0 & 0 & 0 & 1 & 1 & 1 & 0 & 0 & 0 & 1 & 1 & 1 & 0 & 0 & 0 & 1 & 1 & 1 & 0 & 0 & 0 & 1 & 1 & 1 & 0 & 0 & 0 & 1 & 1 & 1 & 0 & 0 & 0 & 1 & 1 & 1\\
0 & 0 & 0 & 0 & 1 & 1 & 0 & 0 & 0 & 0 & 1 & 1 & 0 & 0 & 0 & 0 & 1 & 1 & 0 & 0 & 0 & 0 & 1 & 1 & 0 & 0 & 0 & 0 & 1 & 1 & 0 & 0 & 0 & 0 & 1 & 1\\
0 & 0 & 0 & 0 & 0 & 1 & 0 & 0 & 0 & 0 & 0 & 1 & 0 & 0 & 0 & 0 & 0 & 1 & 0 & 0 & 0 & 0 & 0 & 1 & 0 & 0 & 0 & 0 & 0 & 1 & 0 & 0 & 0 & 0 & 0 & 1
\end{pmatrix}.
\end{eqnarray}
\end{widetext}
For example, the first component in Eq.\ (\ref{YAX}) says that everybody picked up from $O_1$ must go via $O_1O_2$, regardless of whether they are going to $D_1,\cdots,D_6$. So $Y_1=\text{``total number of people on the directed edge $O_1O_2$''}$ is the sum of all the people originating from $O_1$. The second component says that everybody picked up from $O_1$ and $O_2$ must go via $O_2O_3$, etc., until the sixth component, which is the sum of every commuter since they must all pass via $O_6D_1$. Now the 7th component refers to the number of commuters on the directed edge $D_1D_2$. Since everybody who wants to go to $D_1$ has alighted, they do not traverse this edge regardless of which origin they came from. So $Y_7$ excludes $X_{O_1D_1},X_{O_2D_1},X_{O_3D_1},X_{O_4D_1},X_{O_5D_1},X_{O_6D_1}$, and so on for the rest of the components of $\vec{Y}$. This is how the routing matrix $A$ is constructed.


It is obvious in Eq.\ (\ref{YAX}) that there are $36$ unknown components of $\vec{\lambda}=(\lambda_1,\cdots,\lambda_{36})^T$, with $\vec{X}\sim\textrm{Poisson}(\vec{\lambda})$. However, there are only 11 independent equations, corresponding to each of the rows in Eq.\ (\ref{YAX}). This is typical of OD inference problems where the number of ADE that provide measurements is less than the number of OD. Vardi's approach is to figure out the maximum likelihood of $\vec{\lambda}$ from observing $\vec{Y}$. The key idea of overcoming the problem of underdetermination is to make use of moments (as well as the normal approximation given a large sample of measured $\vec{Y}$'s) to generate more and sufficient equations (details in Ref.\ \cite{Vardi96}), leading to an iterative algorithm that would converge to the optimal $\vec{\lambda}$. Let $\vec{\bar{Y}}$ be the mean of all $\vec{Y}$'s which are sampled by measuring the number of commuters at each ADE, and $\vec{S}$ be the corresponding covariance matrix arranged as a column vector. On top of that, let $B$ be a matrix obtained from $A$ where each row of $B$ is the element-wise product of a pair of rows of $A$ (details in Ref.\ \cite{Vardi96}). Note that $\vec{S}$ and $B$ are the result of the normal approximation and the use of first and second order moments to generate more equations. Vardi then constructs the following augmented matrix equation
\[
\begin{pmatrix}\vec{\bar{Y}}\\\vec{S}\end{pmatrix}=
\begin{pmatrix}A\\B\end{pmatrix}\vec{\lambda}.
\] 
This results in an iterative algorithm for computing $\vec{\lambda}$:
\begin{align}\label{lambda}
\lambda_k&\leftarrow\frac{\lambda_k}{\sum_i{a_{ik}}+\sum_i{b_{ik}}}\nonumber\\&\phantom{\leftarrow}\times\left(\sum_i{\frac{a_{ik}\bar{Y}_i}{\sum_p{a_{ip}\lambda_p}}}+\sum_i{\frac{b_{ik}S_i}{\sum_p{b_{ip}\lambda_p}}}\right).
\end{align}
Here, $\lambda_k,\bar{Y}_i,S_i,a_{ik},b_{ik}$ are the components of $\vec{\lambda},\vec{\bar{Y}},\vec{S},A,B$, respectively. Once $\vec{\lambda}$ has been found, then we can normalise with respect to each origin to obtain the probabilities $\zeta_{ij}$.

\subsection{DNN architecture for bus loop service}

For the bus loop service, there are $N$ buses serving the loop of bus stops. Unlike the situation for the general complex networks, here the input data comprises number of people on the bus after the bus leaves a bus stop, $\alpha$; and/or the duration the bus spends stopping at a bus stop, $\beta$. In principle, we can let each input unit be the number of people on each bus, and/or the duration the bus spends stopping at a bus stop as a second feature for the input. Nevertheless, we find that concatenating these time series from all buses into one single long time series as the input turns out to empirically boost performance. This perhaps arises due to complex interactions amongst the buses, viz. the number of people picked up by a leading bus affects the remaining number of people picked up by trailing buses.

Hence for DNNGRU, the input layer comprises $\alpha$ and/or $\beta$, where these features are the concatenation of the time-series of all $N$ buses. If only $\alpha$ or $\beta$ is used, then the input layer comprises that one single unit with a concatenated time-series. The rest is the same as the DNNGRU architecture used for general complex networks, viz. two hidden GRU layers with default activation each with width of 128 units, followed by an output layer with $M_D$ units with a sigmoid activation. So the output is $\zeta_{ij}$ for each destination $j$.

For directly classifying the rank of destination bus stop, the loss function used is sparse categorical cross entropy, which is the standard loss function used in a classification problem with many labels. Then, an argmax is applied to the output layer, such that the unit with the largest value is deemed as the prediction of the destination bus stop of the stipulated rank. This loss function compares the correctness of the predicted destination bus stop with respect to the actual destination bus stop. Direct classification differs from directly regressing for the values of $\zeta_{ij}$ where the loss function used is binary cross entropy that measures the closeness of the values of the predicted $\zeta_{ij}$ from the true value. Finally in the case where dense layers are used in the direct classification, the GRU units are simply replaced by regular densely connected feedforward layers. The input layer is not one single feature, but comprises all values of $\alpha$ which is then fed into two densely connected layers each of width 128 units followed by an output layer with $M_D$ units.

\section{Data Availability}
The data that support the findings of this study are available from the corresponding
author upon reasonable request.

\begin{acknowledgments}
This work was supported by the Joint WASP/NTU Programme (Project No. M4082189). The GPUs at our disposal for this work are RTX 2070 Super, GTX 1650 Super, GTX 1050, GTX 970, Quadro K620. We thank Andri Pradana for letting us use his GTX 970 GPU for training our deep neural networks.
\end{acknowledgments}



\bibliographystyle{naturemag}

\bibliography{Citation}








\end{document}



\title{Inferring origin-destination distribution of agent transfer in a complex network using deep gated recurrent units - Supplemental material}
\author{Vee-Liem Saw}
\email{Vee-Liem@ntu.edu.sg}
\affiliation{Division of Physics and Applied Physics, School of Physical and Mathematical Sciences, Nanyang Technological University, Singapore}
\author{Luca Vismara}
\email{vism0001@e.ntu.edu.sg}
\affiliation{Division of Physics and Applied Physics, School of Physical and Mathematical Sciences, Nanyang Technological University, Singapore}
\author{Suryadi}
\email{sury0013@e.ntu.edu.sg}
\affiliation{Division of Physics and Applied Physics, School of Physical and Mathematical Sciences, Nanyang Technological University, Singapore}
\author{Bo Yang}
\email{yang.bo@ntu.edu.sg}
\affiliation{Division of Physics and Applied Physics, School of Physical and Mathematical Sciences, Nanyang Technological University, Singapore}
\author{Mikael Johansson}
\email{mikaelj@kth.se}
\affiliation{School of Electrical Engineering, KTH Royal Institute of Technology, Stockholm, Sweden}
\author{Lock Yue Chew}
\email{lockyue@ntu.edu.sg}
\affiliation{Division of Physics and Applied Physics, School of Physical and Mathematical Sciences, Nanyang Technological University, Singapore}
%

\date{\today}

%

\maketitle


\section{Linear regression for Vardi's network}

Vardi's network \cite{Vardi96} is defined by the routing matrix $A$, given by:
\begin{align}\label{VardiA}
A&=
\begin{pmatrix}
1 & 0 & 0 & 0 & 0 & 0 & 0 & 0 & 0 & 0 & 0 & 0\\
0 & 1 & 1 & 0 & 0 & 1 & 0 & 0 & 0 & 0 & 0 & 0\\
0 & 0 & 0 & 1 & 0 & 1 & 1 & 0 & 0 &1  & 0 & 0\\
0 & 0 & 0 & 0 & 1 & 0 & 0 & 0 & 0 & 0 & 0 & 0\\
0 & 0 & 0 & 0 & 0 & 0 & 1 & 1 & 0 & 1 & 1 & 0\\
0 & 0 & 1 & 0 & 0 & 1 & 0 & 0 & 1 & 0 & 0 & 0\\
0 & 0 & 0 & 0 & 0 & 0 & 0 & 0 & 0 & 1 & 1 & 1
\end{pmatrix}.
\end{align}
This routing matrix relates the actual number of agents contained in the 12 OD components of $\vec{X}$, and the number of agents at each of the 7 edges in his network $\vec{Y}$, i.e.
\begin{align}\label{YAX}
\vec{Y}&=A\vec{X}.
\end{align}
The 12 components of $\vec{X}$ comprise the number of agents going from $a$ to $b$, $a$ to $c$, $a$ to $d$, $b$ to $a$, $b$ to $c$, $b$ to $d$, $c$ to $a$, $c$ to $b$, $c$ to $d$, $d$ to $a$, $d$ to $b$, $d$ to $c$. The 7 components of $\vec{Y}$ comprise the number of agents on the directed edges $ab$, $ac$, $ba$, $bc$, $cb$, $cd$, $dc$. We denote their components $X_{ab}$ as the number of agents from origin $a$ to destination $b$, and $Y_{ab}$ as the number of agents on the directed edge $ab$, etc. (See \cite{Vardi96}, as these definitions are taken directly from his network given there.)

Unlike our networks where we consider agents propagating in time through the edges sequentially, Vardi views the data as binned over a time interval. In other words, he assumes that over this time interval, if given $\vec{X}$ of all the 12 possible OD, then the number of agents $\vec{Y}$ on the 7 edges over this time interval are given by Eq.\ (\ref{YAX}). This view is in fact also explicitly stated in a Bayesian approach that uses Vardi's network in Ref. \cite{Tebaldi98} (see the abstract of that paper). For an observer, we are only given information of $\vec{Y}$, and the task is to infer the $\vec{X}$ that gave rise to the observation. As there are only 7 equations for 12 unknowns in Eq.\ (\ref{YAX}), this is an underdetermined system.

Vardi's EM algorithm has been elaborated in the Methods section of the main text. Here, we provide the solution by linear regression of the probabilities $\zeta_{ij}$, which is possible if there are additional data: the total number of agents originating from each node, $\vec{V}=(V_a, V_b, V_c, V_d)^T$. This is reasonably obtainable as a counter can track how many agents are leaving each origin node without knowing where they want to go. In the case of a bus system, this is simply tracking how many people are boarding the bus at each bus stop.

With knowledge of $\vec{V}$, we can write $\vec{X}$ in terms of $\vec{V}$ and the probabilities $\zeta_{ij}$. These probabilities $\zeta_{ij}$ are unknown, and now become the variables of interest. From a set of data samples of $\vec{X}$, these cannot be combined as different samples have different sets of OD. Nevertheless under the assumption that the probabilities are constant during the period of interest, the samples can now be combined to regress for this common $\zeta_{ij}$. Eq.\ (\ref{YAX}) becomes:
\begin{align}
Y_{ab}&=\zeta_{ab}V_a\\
Y_{ac}&=(1-\zeta_{ab})V_a+\zeta_{bd}V_b\\
Y_{ba}&=(1-\zeta_{bc})V_b+\zeta_{ca}V_c+\zeta_{da}V_d\\
Y_{bc}&=\zeta_{bc}V_b\\
Y_{cb}&=(1-\zeta_{cd})V_c+(1-\zeta_{dc})V_d\\
Y_{cd}&=\zeta_{ad}V_a+\zeta_{bd}V_b+\zeta_{cd}V_c\\
Y_{dc}&=V_d.
\end{align}
We have used the fact that probabilities originating from the same origin must go to one of the destinations, so that they all sum up to 1. Consequently, instead of 12 components of $\zeta_{ij}$, we only have 8 independent components, viz. $\zeta_{ab},\zeta_{ad},\zeta_{bc},\zeta_{bd},\zeta_{ca},\zeta_{cd},\zeta_{da},\zeta_{dc}$. Note also that the 7th equation $Y_{dc}=V_d$ is not useful, so we only have 6 independent equations.

Given measured data of $\vec{V},\vec{Y}$, these 6 equations can be analytically solved for the 8 independent $\zeta_{ij}$ so that the predicted $\hat{\vec{Y}}$ as compared to the actual $\vec{Y}$ would minimise the mean squared error, analogous to Eq.\ (\ref{LRloss}). We summarise the analytical solution for these optimal $\zeta_{ij}$ as follows. Let $\sigma$ be
\footnotesize
\begin{align}
\sigma&=
\begin{pmatrix}
\displaystyle2\sum_m{V_a^2} & 0 & 0 & \displaystyle-\sum_m{V_aV_b} & 0 & 0 & 0 & 0\\
0 & \displaystyle\sum_m{V_a^2} & 0 & \displaystyle\sum_m{V_aV_b} & 0 & \displaystyle\sum_m{V_aV_c} & 0 & 0\\
0 & 0 & \displaystyle2\sum_m{V_b^2} & 0 & \displaystyle-\sum_m{V_bV_c} & 0 & \displaystyle-\sum_m{V_bV_d} & 0\\
\displaystyle-\sum_m{V_aV_b} & \displaystyle\sum_m{V_aV_b} & 0 & \displaystyle2\sum_m{V_b^2} & 0 & \displaystyle\sum_m{V_bV_c} & 0 & 0\\
0 & 0 & -\displaystyle\sum_m{V_bV_c} & 0 & \displaystyle\sum_m{V_c^2} & 0 & \displaystyle\sum_m{V_cV_d} & 0\\
0 & \displaystyle\sum_m{V_aV_c} & 0 & \displaystyle\sum_m{V_bV_c} & 0 & \displaystyle2\sum_m{V_c^2} & 0 & \displaystyle\sum_m{V_cV_d}\\
0 & 0 & \displaystyle-\sum_m{V_bV_d} & 0 & \displaystyle\sum_m{V_cV_d} & 0 & \displaystyle\sum_m{V_d^2} & 0\\
0 & 0 & 0 & 0 & 0 & \displaystyle\sum_m{V_cV_d} & 0 & \displaystyle\sum_m{V_d^2}
\end{pmatrix}.
\end{align}
\normalsize
The solution is:
\begin{align}
\begin{pmatrix}
\zeta_{ab}\\
\zeta_{ad}\\
\zeta_{bc}\\
\zeta_{bd}\\
\zeta_{ca}\\
\zeta_{cd}\\
\zeta_{da}\\
\zeta_{dc}
\end{pmatrix}&=\sigma^{-1}\cdot
\begin{pmatrix}
\displaystyle\sum_m{V_a(V_a+Y_{ab}-Y_{ac})}\\
\displaystyle\sum_m{V_aY_{cd}}\\
\displaystyle\sum_m{V_b(V_b-Y_{ba}+Y_{bc})}\\
\displaystyle\sum_m{V_b(-V_a+Y_{ac}+Y_{cd})}\\
\displaystyle\sum_m{V_c(-V_b+Y_{ba})}\\
\displaystyle\sum_m{V_c(V_c+V_d-Y_{cb}+Y_{cd})}\\
\displaystyle\sum_m{V_d(-V_b+Y_{ba})}\\
\displaystyle\sum_m{V_d(V_c+V_d-Y_{cb})}
\end{pmatrix}.
\end{align}

\section{Other methods of comparison for Vardi's network: Entropy maximisation and Bayesian inference}

We implement Willumsen's entropy maximisation method to infer the OD trip matrix for Vardi's network \cite{VZ80}. The method is a direct application of the formula given in Ref.\ \cite{VZ80} for entropy maximisation. This turns out to return poor results with coefficient of determination peaking around $r^2\sim0.039$. Clearly, the assumption that the most likely trip matrix would correspond to the one with largest entropy or equivalently, one which minimises the information in the observed data is not necessarily the case with the way agents behave in general networks.

We also carry out a Bayesian inference of OD, whereby given a uniform prior of the parameter $\vec{\lambda}$ and the Poisson likelihood of observing $\vec{Y}$ given $\vec{\lambda}$, we can use Bayes's theorem to obtain the posterior distribution of the parameter $\vec{\lambda}$ given $\vec{Y}$. There exists several Bayesian inference methods which differ in some technical details in the literature \cite{Tebaldi98,Dey94,Carvalho14}. For instance, Ref.\ \cite{Tebaldi98} implemented their Bayesian inference on Vardi's original network in a manner that involves $\vec{X}$, but did not compare with Vardi's EM in terms of performance. Vardi himself also pointed out that their formulation of Bayesian inference is only applicable with one observation of $\vec{Y}$ and not directly generalisable to many samples of $\vec{Y}$. Nevertheless, Vardi did suggest a more direct way of implementing a Bayesian inference (without the need to involve $\vec{X}$) to obtain the posterior distribution of the parameter $\vec{\lambda}$ given $\vec{Y}$, which we adopt here. The details of our implementation are presented in Section \ref{VardiBayes} below. Here, we summarise the results of such a Bayesian inference applied to Vardi's network. It turns out that the model fits the data very poorly: the $r^2$ value is $\sim-0.2$, regardless of the amount of samples of $\vec{Y}$ that are given. Inspection of the fitting of the Bayesian model indicates that underdetermination (there are only 7 equations for 12 unknowns) leads to high uncertainty in the inference. The Bayesian fitting produces a joint posterior distribution for all these 12 unknown components of $\vec{\lambda}$, with rather high degree of spread around the mean values. Hence, the Bayesian model is saying that its posterior prediction of the mean value of $\vec{\lambda}$ is plagued by large uncertainty.

Recall that Vardi's EM creates more equations using moments, to overcome the underdetermination of the equations, whereas our LR introduced in this paper strives to regress for the common coefficients $\zeta_{ij}$ such that all samples of measurements can be combined. Instead, Bayesian inference and entropy maximisation methods do not address the underdetermination problem, and consequently are still burdened by the high uncertainty of their results.


\section{Bayesian inference for Vardi's network} \label{VardiBayes}

To implement a Bayesian inference on Vardi's network, we strive to figure out the probability of the Poisson parameters $\vec{\lambda}$ given the observation of $m$ samples of $\vec{Y}^{(1)},\cdots,\vec{Y}^{(m)}$, i.e. $P(\vec{\lambda}|\vec{Y}^{(1)},\cdots,\vec{Y}^{(m)})$. Using Bayes's theorem, we have
\begin{align}
P(\vec{\lambda}|\vec{Y}^{(1)},\cdots,\vec{Y}^{(m)})&=\frac{P(\vec{Y}^{(1)},\cdots,\vec{Y}^{(m)}|\vec{\lambda})P(\vec{\lambda})}{P(\vec{Y}^{(1)},\cdots,\vec{Y}^{(m)})}.
\end{align}
Here, $\displaystyle P(\vec{\lambda})=\prod_{k=1}^{12}{P(\lambda_k)}$ is the prior for the parameters $\vec{\lambda}$ whose 12 components are the 12 OD Poisson parameters in Vardi's network. Well, recall that in Vardi's network, $\vec{Y}=A\vec{X}$, with $A$ given by Eq.\ (\ref{VardiA}). Also, each component of $\vec{X}$ is a Poisson random variable with corresponding Poisson parameter $\vec{\lambda}$. Thus, each of the 7 components of $\vec{Y}$ is a sum of Poisson random variables, which is just a Poisson random variable with the corresponding sum of the Poisson parameters. This allows a direct fitting of the Bayesian model given the observations of $\vec{Y}$. The model fitting is done using the \emph{pymc3} Python library. In general, such computation is significantly lengthier than Vardi's EM or the training of DNNGRU. As a comparison, training of DNNGRU using 250k datasets takes about 9 hours on an RTX2070 Super GPU, with testing on 50k datasets done within a minute. Vardi's EM takes two days on a CPU to test the same 50k datasets. But the Bayesian fit takes 3 days on a CPU to test merely 1k datasets!

\section{Comparison amongst DNNGRU, LR, EM for the loop}\label{comparemethods}

Figs. 3(g, h, i) in the main text compares the performances amongst DNNGRU, analytically solving the linear regression of data points with respect to Eq.\ (\ref{generalmultivariate}), and Vardi's EM algorithm \cite{Vardi96}. (See Methods for a review of Vardi's EM algorithm and how it compares with LR. See also Section \ref{loop} below for the analytical solution of LR.). This is done for the loop network (without lag, so $l=1$) where Eq.\ (\ref{generalmultivariate}) is directly applicable, and the implementation of Vardi's algorithm is reasonably doable especially since there is a unique path from each origin to each destination. Clearly, DNNGRU is superior to EM which assumes that the OD matrix comprises Poisson random variables and tries to estimate the maximum likelihood that they lead to the observed data on the edges.

The performance of LR fits between DNNGRU and EM. Intriguingly, LR is comparable to EM when $T$ is small (i.e. given a small dataset), but asymptotically matches the performance of DNNGRU for large $T$. Note that DNNGRU is \emph{always} the best, with the exception of one data point where $T=300$ in Fig. 3(g) in the main text when predicting the most popular destination. Whilst DNNGRU can give a prediction even with $T=1$, both LR and EM require $T\geq11$ since the entire loop comprises 11 ADE and these methods require the observation of one full passage from $O_1$ to $D_6$ in order to calculate the necessary quantities. Even so, DNNGRU with $T=1$ is more accurate than LR and EM with $T=20$, thus emphasising its superiority. Figs. 3(h, i) in the main text show the corresponding results for percentage of predictions with $\varepsilon>0.05$ and $1-r^2$, respectively.

\section{Linear regression for a bus loop system}\label{loop}

We consider a bus loop system where $N$ buses pick up people from $M_O$ origin bus stops and then let them alight at $M_D$ destination bus stops (c.f. Fig.\ 1(d) in the main text). For the case of a complex network where the network is a directed loop (c.f. Fig.\ 4(a) in the main text), the results here for regular buses (but not semi-express buses as the network topology is different) are applicable, though we need to obtain the number of agents spawning from an origin node as well as the number of agents arriving at a destination node as the difference between two adjacent nodes.

The multivariate linear system governing the number of people boarding $x_i$ and alighting $y_j$ are:
\begin{align}\label{generalmultivariate}
y_j=\sum_{i=1}^{M_O}{\zeta_{ij}x_i},\textrm{ for }j=1,\cdots,M_D,
\end{align}
subject to the constraints:
\begin{align}\label{constraints}
\sum_{j=1}^{M_D}{\zeta_{ij}}=1,\textrm{ for }i=1,\cdots,M_O.
\end{align}

Note that since the $M_O$ origins are all before the $M_D$ destinations in the loop, the number of people alighting at the last destination is the sum of everybody who boarded minus the sum of everybody who alighted before the last destination:
\begin{align}\label{lastalight}
y_{M_D}=x_1+\cdots+x_{M_O}-y_1-\cdots-y_{M_D-1}.
\end{align}
With this and the constraints Eq.\ (\ref{constraints}), we can show that  the last component of Eq.\ (\ref{generalmultivariate}) (where $j=M_D$) is not independent from all the other components. For instance, consider the right-hand side:
\begin{align}
\sum_{i=1}^{M_O}{\zeta_{i{M_D}}x_i}&=\sum_{i=1}^{M_O}{\left(1-\sum_{j=1}^{M_D-1}{\zeta_{ij}}\right)x_i}\text{ using the constraints Eq.\ (\ref{constraints})}\\
&=\sum_{i=1}^{M_O}{x_i}-\sum_{j=1}^{M_D-1}{\left(\sum_{i=1}^{M_O}{\zeta_{ij}x_i}\right)}\\
&=\sum_{i=1}^{M_O}{x_i}-\sum_{j=1}^{M_D-1}{y_j}\text{ using Eq.\ (\ref{generalmultivariate})}\\
&=y_{M_D}\textrm{ using Eq.\ (\ref{lastalight})},
\end{align}
which is equal to the left-hand side. Thus, only $M_D-1$ equations in Eq.\ (\ref{generalmultivariate}) are linearly independent.

Eq.\ (\ref{generalmultivariate}) is satisfied for every carrier of the agent transfer linking the nodes in a complex network, or bus going across bus stops. By linear regression, the dataset $(x_i,y_j)$ is fitted so that the predicted $\hat{y}_j$ according to Eq.\ (\ref{generalmultivariate}) as compared to the actual $y_j$ would minimise the mean squared error:
\begin{align}\label{LRloss}
J=\sum_{m,j}{(\hat{y}_j-y_j)^2},
\end{align}
where $m$ is the number of data points $(x_i,y_j)$. This is a convex minimisation problem, so a global minimum of $J$ exists, which can be analytically calculated. This involves setting $\displaystyle\frac{\partial J}{\partial\zeta_{ij}}=0$, leading to a system of linear equations for all independent components of $\zeta_{ij}$. We summarise the solutions for regular and semi-express bus systems with $N=M_O=M_D=6$ in the following subsections.

\subsection{Regular buses}

For regular buses, every bus would board and alight commuters at every bus stop. Therefore, every bus satisfies Eq.\ (\ref{generalmultivariate}). Let $\sigma$ be an $M_O$ by $M_O$ matrix, with components:
\begin{align}
\sigma_{ij}&=\sum_m{x_ix_j}.
\end{align}
This leads to the following analytical solution for $\zeta_{ij}$ which minimises $J$. For $N=M_O=M_D=6$, the solution is:
\begin{align}
\begin{pmatrix}
\zeta_{1j}\\
\zeta_{2j}\\
\zeta_{3j}\\
\zeta_{4j}\\
\zeta_{5j}\\
\zeta_{6j}
\end{pmatrix}&=\sigma^{-1}\cdot
\begin{pmatrix}
\displaystyle\sum_m{x_1y_j}\\
\displaystyle\sum_m{x_2y_j}\\
\displaystyle\sum_m{x_3y_j}\\
\displaystyle\sum_m{x_4y_j}\\
\displaystyle\sum_m{x_5y_j}\\
\displaystyle\sum_m{x_6y_j},
\end{pmatrix}
\end{align}
where $j=1,\cdots,M_D-1$. The $j=M_D$ components are obtained using the constraints Eq.\ (\ref{constraints}).

\subsection{Semi-express buses}

For semi-express buses, different buses serve different subsets of origin bus stops. Therefore, different buses satisfy different versions of Eq.\ (\ref{generalmultivariate}) and we need to work them out specifically. For $N=M_O=M_D=6$, let the $N=6$ buses be $A,B,C,D,E,F$. Each bus picks up from three origins, so suppose $A$ picks up from origins $1,2,3$ (so $x_{A4}=x_{A5}=x_{A6}=0$); $B$ picks up from origins $2,3,4$ (so $x_{B1}=x_{B5}=x_{B6}=0$); $C$ picks up from origins $3,4,5$ (so $x_{C1}=x_{C2}=x_{C6}=0$); $D$ picks up from origins $4,5,6$ (so $x_{D1}=x_{D2}=x_{D3}=0$); $E$ picks up from origins $1,5,6$ (so $x_{E2}=x_{E3}=x_{E4}=0$); $F$ picks up from origins $1,2,6$ (so $x_{F3}=x_{F4}=x_{F5}=0$). Then, we have the following equations for the semi-express buses $A,B,C,D,E,F$:
\begin{align}
y_{Aj}&=\zeta_{1j}x_{A1}+\zeta_{2j}x_{A2}+\zeta_{3j}x_{A3}\\
y_{Bj}&=\zeta_{2j}x_{B2}+\zeta_{j3}x_{B3}+\zeta_{4j}x_{B4}\\
y_{Cj}&=\zeta_{3j}x_{C3}+\zeta_{4j}x_{C4}+\zeta_{5j}x_{C5}\\
y_{Dj}&=\zeta_{4j}x_{D4}+\zeta_{5j}x_{D5}+\zeta_{6j}x_{D6}\\
y_{Ej}&=\zeta_{1j}x_{E1}+\zeta_{5j}x_{E5}+\zeta_{6j}x_{E6}\\
y_{Fj}&=\zeta_{1j}x_{F1}+\zeta_{2j}x_{F2}+\zeta_{6j}x_{F6},
\end{align}
where $j=1,\cdots,M_D-1$. Let $\sigma_A,\sigma_B,\sigma_C,\sigma_D,\sigma_E,\sigma_F$ be $M_O$ by $M_O$ matrices, with components:
\begin{align}
\sigma_{Aij}&=\sum_m{x_{Ai}x_{Aj}}\\
\sigma_{Bij}&=\sum_m{x_{Bi}x_{Bj}}\\
\sigma_{Cij}&=\sum_m{x_{Ci}x_{Cj}}\\
\sigma_{Dij}&=\sum_m{x_{Di}x_{Dj}}\\
\sigma_{Eij}&=\sum_m{x_{Ei}x_{Ej}}\\
\sigma_{Fij}&=\sum_m{x_{Fi}x_{Fj}},
\end{align}
and
\begin{align}
\sigma_{\textrm{ semi-express}}&=\sigma_{A}+\sigma_{B}+\sigma_{C}+\sigma_{D}+\sigma_{E}+\sigma_{F}.
\end{align}
The solution is:
\begin{align}
\begin{pmatrix}
\zeta_{1j}\\
\zeta_{2j}\\
\zeta_{3j}\\
\zeta_{4j}\\
\zeta_{5j}\\
\zeta_{6j}
\end{pmatrix}&=\sigma_{\textrm{ semi-express}}^{-1}\cdot
\begin{pmatrix}
\displaystyle\sum_m{x_{A1}y_{Aj}+x_{E1}y_{Ej}+x_{F1}y_{Fj}}\\
\displaystyle\sum_m{x_{A2}y_{Aj}+x_{B2}y_{Bj}+x_{F2}y_{Fj}}\\
\displaystyle\sum_m{x_{A3}y_{Aj}+x_{B3}y_{Bj}+x_{C3}y_{Cj}}\\
\displaystyle\sum_m{x_{B4}y_{Bj}+x_{C4}y_{Cj}+x_{D4}y_{Dj}}\\
\displaystyle\sum_m{x_{C5}y_{Cj}+x_{D5}y_{Dj}+x_{E5}y_{Ej}}\\
\displaystyle\sum_m{x_{D6}y_{Dj}+x_{E6}y_{Ej}+x_{F6}y_{Fj}}
\end{pmatrix}
\end{align}
where $j=1,\cdots,M_D-1$. The $j=M_D$ components are obtained using the constraints Eq.\ (\ref{constraints}).

\section{Analytical treatment for lattice with \texorpdfstring{$M_O=M_D=3$}{MO=MD=3}}

For this lattice, here are the shortest paths from origin nodes $O_1,O_2,O_3$ to $D_1,D_2,D_3$, respectively (c.f. Fig.\ 4(g) in the main text):
\begin{align}
O_1&\rightarrow
\begin{cases}
O_2\rightarrow D_1\textrm{ or}\\
O_3\rightarrow D_1\textrm{ or}\\
D_2\rightarrow D_1\textrm{ or}\\
D_3\rightarrow D_1,
\end{cases}\\
O_1&\rightarrow D_2,\\
O_1&\rightarrow D_3,\\
O_2&\rightarrow D_1,\\
O_2&\rightarrow
\begin{cases}
O_3\rightarrow D_2\textrm{ or}\\
O_1\rightarrow D_2\textrm{ or}\\
D_3\rightarrow D_2\textrm{ or}\\
D_1\rightarrow D_2,
\end{cases}\\
O_2&\rightarrow D_3,\\
O_3&\rightarrow D_1,\\
O_3&\rightarrow D_2,\\
O_3&\rightarrow
\begin{cases}
O_1\rightarrow D_3\textrm{ or}\\
O_2\rightarrow D_3\textrm{ or}\\
D_1\rightarrow D_3\textrm{ or}\\
D_2\rightarrow D_3.
\end{cases}
\end{align}

Let $O_1D_2$ denote the number of agents on the directed edge from $O_1$ to $D_2$, and similarly for other directed edges. By inspection, we can write down this equation for the number of agents on the directed edge $O_1D_2$:
\begin{align}\label{O1D2}
O_1D_2&=\zeta_{12}X_1+\frac{1}{4}\zeta_{11}X_1+O_2O_1(-1)\\
&=\zeta_{12}X_1+D_2D_1(+1)+O_2O_1(-1).
\end{align}
Here, $X_1$ is the number of agents originating from the origin node $O_1$ at this time step, where $X_1$ is not known (and is varying over time). The number of agents from $O_1$ going to $D_2$ is given by $\zeta_{12}X_1$, since $\zeta_{12}$ is the probability for an agent from $O_1$ going to $D_2$. This is the first term on the right-hand side of Eq.\ (\ref{O1D2}). The second term $\zeta_{11}X_1/4$ is due to the agents from $O_1$ having probability $\zeta_{11}$ of going to destination $D_1$. Amongst these agents, $1/4$ of them (on average) would take the path $O_1\rightarrow D_2\rightarrow D_1$, since one of the four shortest paths are selected randomly. Incidentally, this term $\zeta_{11}X_1/4$ is equal to $D_2D_1(+1)$ where ``$(+1)$'' implies ``at the next time step'' because all agents would move to $D_2D_1$ at the next time step. The third term $O_2O_1(-1)$ is due to agents from $O_2O_1$ from the previous time step now moving on to $O_1D_2$, where ``$(-1)$'' implies ``from the previous time step''.

Similarly, we obtain five more such equations:
\begin{align}
O_1D_3&=\zeta_{13}X_1+\frac{1}{4}\zeta_{11}X_1+O_3O_1(-1)\\
&=\zeta_{13}X_1+D_3D_1(+1)+O_3O_1(-1)\\
O_2D_1&=\zeta_{21}X_2+\frac{1}{4}\zeta_{22}X_2+O_1O_2(-1)\\
&=\zeta_{21}X_2+D_1D_2(+1)+O_1O_2(-1)\\
O_2D_3&=\zeta_{23}X_2+\frac{1}{4}\zeta_{22}X_2+O_3O_2(-1)\\
&=\zeta_{23}X_2+D_3D_2(+1)+O_3O_2(-1)\\
O_3D_1&=\zeta_{31}X_3+\frac{1}{4}\zeta_{33}X_3+O_1O_3(-1)\\
&=\zeta_{31}X_3+D_1D_3(+1)+O_1O_3(-1)\\
O_3D_2&=\zeta_{32}X_3+\frac{1}{4}\zeta_{33}X_3+O_2O_3(-1)\\
&=\zeta_{32}X_3+D_2D_3(+1)+O_2O_3(-1).
\end{align}
Therefore, $\zeta_{12}X_1,\zeta_{13}X_1,\zeta_{21}X_2,\zeta_{23}X_2,\zeta_{31}X_3,\zeta_{32}X_3$ are given by the averages:
\begin{align}
\zeta_{12}X_1&=O_1D_2-D_2D_1(+1)-O_2O_1(-1)\label{Zeta1}\\
\zeta_{13}X_1&=O_1D_3-D_3D_1(+1)-O_3O_1(-1)\\
\zeta_{21}X_2&=O_2D_1-D_1D_2(+1)-O_1O_2(-1)\\
\zeta_{23}X_2&=O_2D_3-D_3D_2(+1)-O_3O_2(-1)\\
\zeta_{31}X_3&=O_3D_1-D_1D_3(+1)-O_1O_3(-1)\\
\zeta_{32}X_3&=O_3D_2-D_2D_3(+1)-O_2O_3(-1),
\end{align}
where the averages are over the measured samples of number of agents on the respective directed edges at each time step.

Incidentally by inspection, we see that for $\zeta_{11}X_1$, $\zeta_{22}X_2$ and $\zeta_{33}X_3$, these are given by the averages:
\begin{align}
\zeta_{11}X_1&=D_2D_1(+1)+D_3D_1(+1) + O_1O_2 + O_1O_3\\
\zeta_{22}X_2&=D_1D_2(+1)+D_3D_2(+1) + O_2O_3 + O_2O_1\\
\zeta_{33}X_3&=D_1D_3(+1)+D_2D_3(+1) + O_3O_1 + O_3O_2,\label{Zeta6}
\end{align}
where the averages are over the measured samples of number of agents on the respective directed edges at each time step. Each of the respective four terms corresponds to each of the four possible paths that can be taken.

Recall that $X_1,X_2,X_3$ are unknown and varying in time. Nevertheless, we can obtain all the $\zeta_{ij}$ by using the constraints Eq.\ (\ref{constraints}):
\begin{align}\label{Xi}
\sum_{j=1}^3{\zeta_{ij}X_i}=X_i,
\end{align}
for $i=1,2, 3$. Therefore,
\begin{align}
\zeta_{ij}&=\frac{\zeta_{ij}X_i}{\sum\limits_{k=1}^3{\zeta_{ik}X_i}}.
\end{align}
Note that the expression for $X_i$ in Eq.\ (\ref{Xi}) is the \emph{average} of the samples. In fact, what we actually deduce from the edges via Eqs.\ (\ref{Zeta1})-(\ref{Zeta6}) are these quantities $Z_{ij}=\zeta_{ij}X_i$, which allow us to get $\zeta_{ij}$ thanks to the constraint Eq.\ (\ref{constraints}). Writing in terms of $Z_{ij}$, the probabilities $\zeta_{ij}$ are:
\begin{align}
\zeta_{ij}&=\frac{Z_{ij}}{\sum\limits_{k=1}^{3}{Z_{ik}}},
\end{align}
where we do not need to involve the unknowns $X_i$ at all.

\section{Details of the lattice, random, and small-world networks}

For the lattice, random and small-world networks, respectively, as displayed in Figs.\ 4(b-d) of the main text, we provide the adjacency matrices, the set of shortest paths, the set of active directed edges (ADE, i.e. the edges that are actually being traversed as part of the shortest paths from $O_i$ to $D_j$), as well as the set of edges used for each origin node to get to all destinations. These three networks have the same number of 12 nodes (6 origins, 6 destinations) and same number of 24 edges (or 48 directed edges). They differ in their topologies, and provide a comparison for how the resulting OD inference properties turn out. With these, the number of overlapping edges between pairs of origin nodes to get to any destinations can be computed, as presented in Table 1 of the main text.

\subsection{Lattice}

The adjacency matrix for the lattice $A_L$ is:
\begin{align}
A_L&=
\begin{pmatrix}
0 & 1 & 1 & 0 & 0 & 0 & 0 & 0 & 0 & 0 & 1 & 1\\
1 & 0 & 1 & 1 & 0 & 0 & 0 & 0 & 0 & 0 & 0 & 1\\
1 & 1 & 0 & 1 & 1 & 0 & 0 & 0 & 0 & 0 & 0 & 0\\
0 & 1 & 1 & 0 & 1 & 1 & 0 & 0 & 0 & 0 & 0 & 0\\
0 & 0 & 1 & 1 & 0 & 1 & 1 & 0 & 0 & 0 & 0 & 0\\
0 & 0 & 0 & 1 & 1 & 0 & 1 & 1 & 0 & 0 & 0 & 0\\
0 & 0 & 0 & 0 & 1 & 1 & 0 & 1 & 1 & 0 & 0 & 0\\
0 & 0 & 0 & 0 & 0 & 1 & 1 & 0 & 1 & 1 & 0 & 0\\
0 & 0 & 0 & 0 & 0 & 0 & 1 & 1 & 0 & 1 & 1 & 0\\
0 & 0 & 0 & 0 & 0 & 0 & 0 & 1 & 1 & 0 & 1 & 1\\
1 & 0 & 0 & 0 & 0 & 0 & 0 & 0 & 1 & 1 & 0 & 1\\
1 & 1 & 0 & 0 & 0 & 0 & 0 & 0 & 0 & 1 & 1 & 0
\end{pmatrix}.
\end{align}

Here is the set of shortest paths from $O_i$ to $D_j$ for the lattice:
\begin{align}
O_1&\rightarrow
\begin{cases}
O_3\rightarrow O_5\rightarrow D_1\textrm{ or}\\
D_5\rightarrow D_3\rightarrow D_1,
\end{cases}\\
O_1&\rightarrow
\begin{cases}
D_5\rightarrow D_3\rightarrow D_2\textrm{ or}\\
D_5\rightarrow D_4\rightarrow D_2\textrm{ or}\\
D_6\rightarrow D_4\rightarrow D_2,
\end{cases}\\
O_1&\rightarrow D_5\rightarrow D_3\\
O_1&\rightarrow
\begin{cases}
D_5\rightarrow D_4\textrm{ or}\\
D_6\rightarrow D_4,
\end{cases}\\
O_1&\rightarrow D_5,\\
O_1&\rightarrow D_6,\\
O_2&\rightarrow
\begin{cases}
O_4\rightarrow O_6\rightarrow D_1\textrm{ or}\\
O_4\rightarrow O_5\rightarrow D_1\textrm{ or}\\
O_3\rightarrow O_5\rightarrow D_1,
\end{cases}\\
O_2&\rightarrow
\begin{cases}
O_4\rightarrow O_6\rightarrow D_2\textrm{ or}\\
D_6\rightarrow D_4\rightarrow D_2,
\end{cases}\\
O_2&\rightarrow
\begin{cases}
D_6\rightarrow D_4\rightarrow D_3\textrm{ or}\\
D_6\rightarrow D_5\rightarrow D_3\textrm{ or}\\
O_1\rightarrow D_5\rightarrow D_3,
\end{cases}\\
O_2&\rightarrow D_6\rightarrow D_4,\\
O_2&\rightarrow
\begin{cases}
D_6\rightarrow D_5\textrm{ or}\\
O_1\rightarrow D_5,
\end{cases}\\
O_2&\rightarrow D_6,\\
O_3&\rightarrow O_5\rightarrow D_1,\\
O_3&\rightarrow
\begin{cases}
O_5\rightarrow D_1\rightarrow D_2\textrm{ or}\\
O_5\rightarrow O_6\rightarrow D_2\textrm{ or}\\
O_4\rightarrow O_6\rightarrow D_2,
\end{cases}\\
O_3&\rightarrow
\begin{cases}
O_5\rightarrow D_1\rightarrow D_3\textrm{ or}\\
O_1\rightarrow D_5\rightarrow D_3,
\end{cases}\\
O_3&\rightarrow
\begin{cases}
O_1\rightarrow D_5\rightarrow D_4\textrm{ or}\\
O_1\rightarrow D_6\rightarrow D_4\textrm{ or}\\
O_2\rightarrow D_6\rightarrow D_4,
\end{cases}\\
O_3&\rightarrow O_1\rightarrow D_5,\\
O_3&\rightarrow
\begin{cases}
O_1\rightarrow D_6\textrm{ or}\\
O_2\rightarrow D_6,
\end{cases}\\
O_4&\rightarrow
\begin{cases}
O_6\rightarrow D_1\textrm{ or}\\
O_5\rightarrow D_1,
\end{cases}\\
O_4&\rightarrow O_6\rightarrow D_2,\\
O_4&\rightarrow
\begin{cases}
O_6\rightarrow D_2\rightarrow D_3\textrm{ or}\\
O_6\rightarrow D_1\rightarrow D_3\textrm{ or}\\
O_5\rightarrow D_1\rightarrow D_3,
\end{cases}\\
O_4&\rightarrow
\begin{cases}
O_6\rightarrow D_2\rightarrow D_4\textrm{ or}\\
O_2\rightarrow D_6\rightarrow D_4,
\end{cases}\\
O_4&\rightarrow
\begin{cases}
O_2\rightarrow D_6\rightarrow D_5\textrm{ or}\\
O_2\rightarrow O_1\rightarrow D_5\textrm{ or}\\
O_3\rightarrow O_1\rightarrow D_5,
\end{cases}\\
O_4&\rightarrow O_2\rightarrow D_6,\\
O_5&\rightarrow D_1,\\
O_5&\rightarrow
\begin{cases}
D_1\rightarrow D_2\textrm{ or}\\
O_6\rightarrow D_2,
\end{cases}\\
O_5&\rightarrow D_1\rightarrow D_3,\\
O_5&\rightarrow
\begin{cases}
D_1\rightarrow D_3\rightarrow D_4\textrm{ or}\\
D_1\rightarrow D_2\rightarrow D_4\textrm{ or}\\
O_6\rightarrow D_2\rightarrow D_4,
\end{cases}\\
O_5&\rightarrow
\begin{cases}
O_3\rightarrow O_1\rightarrow D_5\textrm{ or}\\
D_1\rightarrow D_3\rightarrow D_5,
\end{cases}\\
O_5&\rightarrow
\begin{cases}
O_3\rightarrow O_1\rightarrow D_6\textrm{ or}\\
O_3\rightarrow O_2\rightarrow D_6\textrm{ or}\\
O_4\rightarrow O_2\rightarrow D_6,
\end{cases}\\
O_6&\rightarrow D_1,\\
O_6&\rightarrow D_2,\\
O_6&\rightarrow
\begin{cases}
D_2\rightarrow D_3\textrm{ or}\\
D_1\rightarrow D_3,
\end{cases}\\
O_6&\rightarrow D_2\rightarrow D_4\\
O_6&\rightarrow
\begin{cases}
D_2\rightarrow D_4\rightarrow D_5\textrm{ or}\\
D_2\rightarrow D_3\rightarrow D_5\textrm{ or}\\
D_1\rightarrow D_3\rightarrow D_5,
\end{cases}\\
O_6&\rightarrow
\begin{cases}
O_4\rightarrow O_2\rightarrow D_6\textrm{ or}\\
D_2\rightarrow D_4\rightarrow D_6.
\end{cases}
\end{align}

The set of 38 ADE for the lattice is:
\begin{gather}
O_1O_3,O_2O_3,O_2O_4,O_3O_4,O_3O_5,O_4O_5,O_4O_6,O_5O_6,\\
O_5D_1,O_6D_1,O_6D_2,D_1D_2,D_1D_3,D_2D_3,D_2D_4,D_3D_4,\\
D_3D_5,D_4D_5,D_4D_6,O_2O_1,O_3O_1,O_3O_2,O_4O_2,O_4O_3,\\
O_5O_3,O_5O_4,O_6O_4,D_3D_1,D_3D_2,D_4D_2,D_4D_3,D_5D_3,\\
D_5D_4,D_6D_4,D_6D_5,O_1D_5,O_1D_6,O_2D_6.
\end{gather}

Here are the edges that are being traversed by each origin node $O_i$ in the lattice, to get to any destination:
\begin{align}
O_1&=O_1O_3,O_1D_5,O_1D_6,O_3O_5,O_5D_1,D_5D_3,D_3D_1,D_3D_2,\nonumber\\
&\phantom{=}D_5D_4,D_4D_2,D_6D_4\\
O_2&=O_2O_4,O_2O_3,O_2D_6,O_2O_1,O_4O_6,O_6D_1,O_4O_5,O_5D_1,\nonumber\\
&\phantom{=}O_3O_5,O_6D_2,D_6D_4,D_4D_2,D_4D_3,D_6D_5,D_5D_3,O_1D_5\\
O_3&=O_3O_5,O_3O_4,O_3O_1,O_3O_2,O_5D_1,D_1D_2,O_5O_6,O_6D_2,\nonumber\\
&\phantom{=}O_4O_6,D_1D_3,O_1D_5,D_5D_3,D_5D_4,O_1D_6,D_6D_4,O_2D_6\\
O_4&=O_4O_6,O_4O_5,O_4O_2,O_4O_3,O_6D_1,O_5D_1,O_6D_2,D_2D_3,\nonumber\\
&\phantom{=}D_1D_3,D_2D_4,O_2D_6,D_6D_4,D_6D_5,O_2O_1,O_1D_5,O_3O_1\\
O_5&=O_5D_1,O_5O_6,O_5O_3,O_5O_4,D_1D_2,O_6D_2,D_1D_3,D_3D_4,\nonumber\\
&\phantom{=}D_2D_4,O_3O_1,O_1D_5,D_3D_5,O_1D_6,O_3O_2,O_2D_6,O_4O_2\\
O_6&=O_6D_1,O_6D_2,O_6O_4,D_2D_3,D_1D_3,D_2D_4,D_4D_5,D_3D_5,\nonumber\\
&\phantom{=}O_4O_2,O_2D_6,D_4D_6.
\end{align}

\subsection{Random network}

The adjacency matrix for the random network $A_R$ is:
\begin{align}
A_R&=
\begin{pmatrix}
0 & 1 & 0 & 1 & 0 & 1 & 1 & 0 & 1 & 0 & 0 & 1\\
1 & 0 & 1 & 0 & 0 & 1 & 0 & 1 & 1 & 1 & 0 & 1\\
0 & 1 & 0 & 0 & 0 & 0 & 0 & 1 & 0 & 0 & 0 & 0\\
1 & 0 & 0 & 0 & 1 & 0 & 1 & 1 & 0 & 0 & 0 & 1\\
0 & 0 & 0 & 1 & 0 & 0 & 0 & 0 & 1 & 0 & 1 & 0\\
1 & 1 & 0 & 0 & 0 & 0 & 0 & 1 & 0 & 0 & 0 & 0\\
1 & 0 & 0 & 1 & 0 & 0 & 0 & 0 & 0 & 0 & 0 & 0\\
0 & 1 & 1 & 1 & 0 & 1 & 0 & 0 & 0 & 0 & 1 & 0\\
1 & 1 & 0 & 0 & 1 & 0 & 0 & 0 & 0 & 1 & 0 & 1\\
0 & 1 & 0 & 0 & 0 & 0 & 0 & 0 & 1 & 0 & 0 & 0\\
0 & 0 & 0 & 0 & 1 & 0 & 0 & 1 & 0 & 0 & 0 & 1\\
1 & 1 & 0 & 1 & 0 & 0 & 0 & 0 & 1 & 0 & 1 & 0
\end{pmatrix}.
\end{align}

Here is the set of shortest paths from $O_i$ to $D_j$ for the random network:
\begin{align}
O_1&\rightarrow D_1,\\
O_1&\rightarrow
\begin{cases}
O_6\rightarrow D_2\textrm{ or}\\
O_2\rightarrow D_2\textrm{ or}\\
O_4\rightarrow D_2,
\end{cases}\\
O_1&\rightarrow D_3\\
O_1&\rightarrow
\begin{cases}
O_2\rightarrow D_4\textrm{ or}\\
D_3\rightarrow D_4,
\end{cases}\\
O_1&\rightarrow D_6\rightarrow D_5,\\
O_1&\rightarrow D_6,\\
O_2&\rightarrow O_1\rightarrow D_1,\\
O_2&\rightarrow D_2,\\
O_2&\rightarrow D_3,\\
O_2&\rightarrow D_4,\\
O_2&\rightarrow
\begin{cases}
D_2\rightarrow D_5\textrm{ or}\\
D_6\rightarrow D_5,
\end{cases}\\
O_2&\rightarrow D_6,\\
O_3&\rightarrow
\begin{cases}
O_2\rightarrow O_1\rightarrow D_1\textrm{ or}\\
D_2\rightarrow O_4\rightarrow D_1,
\end{cases}\\
O_3&\rightarrow D_2,\\
O_3&\rightarrow O_2\rightarrow D_3,\\
O_3&\rightarrow O_2\rightarrow D_4,\\
O_3&\rightarrow D_2\rightarrow D_5,\\
O_3&\rightarrow O_2\rightarrow D_6,\\
O_4&\rightarrow D_1,\\
O_4&\rightarrow D_2,\\
O_4&\rightarrow
\begin{cases}
O_1\rightarrow D_3\textrm{ or}\\
O_5\rightarrow D_3\textrm{ or}\\
D_6\rightarrow D_3,
\end{cases}\\
O_4&\rightarrow
\begin{cases}
O_1\rightarrow D_3\rightarrow D_4\textrm{ or}\\
O_5\rightarrow D_3\rightarrow D_4\textrm{ or}\\
D_2\rightarrow O_2\rightarrow D_4\textrm{ or}\\
D_6\rightarrow D_3\rightarrow D_4 \textrm{ (capped to 4 choices)},
\end{cases}\\
O_4&\rightarrow
\begin{cases}
O_5\rightarrow D_5\textrm{ or}\\
D_2\rightarrow D_5\textrm{ or}\\
D_6\rightarrow D_5,
\end{cases}\\
O_4&\rightarrow D_6,\\
O_5&\rightarrow O_4\rightarrow D_1,\\
O_5&\rightarrow O_4\rightarrow D_2,\\
O_5&\rightarrow D_3,\\
O_5&\rightarrow D_3\rightarrow D_4,\\
O_5&\rightarrow D_5,\\
O_5&\rightarrow
\begin{cases}
O_4\rightarrow D_6\textrm{ or}\\
D_3\rightarrow D_6\textrm{ or}\\
D_5\rightarrow D_6,
\end{cases}\\
O_6&\rightarrow O_1\rightarrow D_1,\\
O_6&\rightarrow D_2,\\
O_6&\rightarrow
\begin{cases}
O_1\rightarrow D_3\textrm{ or}\\
O_2\rightarrow D_3,
\end{cases}\\
O_6&\rightarrow O_2\rightarrow D_4\\
O_6&\rightarrow D_2\rightarrow D_5\\
O_6&\rightarrow
\begin{cases}
O_1\rightarrow D_6\textrm{ or}\\
O_2\rightarrow D_6.
\end{cases}
\end{align}

The set of 32 ADE for the random network is:
\begin{gather}
O_1O_2,O_1O_4,O_1O_6,O_1D_1,O_1D_3,O_1D_6,O_2D_2,O_2D_3,\\
O_2D_4,O_2D_6,O_3D_2,O_4O_5,O_4D_1,O_4D_2,O_4D_6,O_5D_3,\\
O_5D_5,O_6D_2,D_3D_4,D_3D_6,D_5D_6,D_2D_5,O_2O_1,O_4O_1,\\
O_6O_1,O_3O_2,O_6O_2,D_2O_2,O_5O_4,D_2O_4,D_6D_3,D_6D_5.
\end{gather}

Here are the edges that are being traversed by each origin node $O_i$ in the random network, to get to any destination:
\begin{align}
O_1&=O_1D_1,O_1O_6,O_1O_2,O_1O_4,O_1D_3,O_1D_6,O_6D_2,O_2D_2,\nonumber\\
&\phantom{=}O_4D_2,O_2D_4,D_3D_4,D_6D_5\\
O_2&=O_2O_1,O_2D_2,O_2D_3,O_2D_4,O_2D_6,O_1D_1,D_2D_5,D_6D_5\\
O_3&=O_3O_2,O_3D_2,O_2O_1,O_1D_1,D_2O_4,O_4D_1,O_2D_3,O_2D_4,\nonumber\\
&\phantom{=}D_2D_5,O_2D_6\\
O_4&=O_4D_1,O_4D_2,O_4O_1,O_4O_5,O_4D_6,O_1D_3,O_5D_3,D_6D_3,\nonumber\\
&\phantom{=}D_3D_4,D_2O_2,O_2D_4,O_5D_5,D_2D_5,D_6D_5\\
O_5&=O_5O_4,O_5D_3,O_5D_5,O_4D_1,O_4D_2,D_3D_4,O_4D_6,D_3D_6,\nonumber\\
&\phantom{=}D_5D_6\\
O_6&=O_6O_1,O_6D_2,O_6O_2,O_1D_1,O_1D_3,O_2D_3,O_2D_4,D_2D_5,\nonumber\\
&\phantom{=}O_1D_6,O_2D_6.
\end{align}

\subsection{Small-world network}

The adjacency matrix for the small-world network $A_{SW}$ is:

\begin{align}
A_{SW}&=
\begin{pmatrix}
0 & 1 & 1 & 1 & 0 & 1 & 0 & 1 & 0 & 1 & 1 & 1\\
1 & 0 & 1 & 1 & 0 & 0 & 0 & 0 & 0 & 0 & 0 & 0\\
1 & 1 & 0 & 1 & 1 & 0 & 0 & 0 & 0 & 0 & 0 & 0\\
1 & 1 & 1 & 0 & 1 & 0 & 0 & 1 & 0 & 0 & 0 & 0\\
0 & 0 & 1 & 1 & 0 & 1 & 0 & 0 & 0 & 0 & 0 & 0\\
1 & 0 & 0 & 0 & 1 & 0 & 1 & 0 & 0 & 0 & 0 & 0\\
0 & 0 & 0 & 0 & 0 & 1 & 0 & 1 & 1 & 0 & 0 & 0\\
1 & 0 & 0 & 1 & 0 & 0 & 1 & 0 & 1 & 1 & 1 & 0\\
0 & 0 & 0 & 0 & 0 & 0 & 1 & 1 & 0 & 1 & 0 & 0\\
1 & 0 & 0 & 0 & 0 & 0 & 0 & 1 & 1 & 0 & 1 & 0\\
1 & 0 & 0 & 0 & 0 & 0 & 0 & 1 & 0 & 1 & 0 & 1\\
1 & 0 & 0 & 0 & 0 & 0 & 0 & 0 & 0 & 0 & 1 & 0
\end{pmatrix}.
\end{align}

Here is the set of shortest paths from $O_i$ to $D_j$ for the small-world network:
\begin{align}
O_1&\rightarrow
\begin{cases}
O_6\rightarrow D_1\textrm{ or}\\
D_2\rightarrow D_1,
\end{cases}\\
O_1&\rightarrow D_2,\\
O_1&\rightarrow
\begin{cases}
D_2\rightarrow D_3\textrm{ or}\\
D_4\rightarrow D_3,
\end{cases}\\
O_1&\rightarrow D_4\\
O_1&\rightarrow D_5,\\
O_1&\rightarrow D_6,\\
O_2&\rightarrow
\begin{cases}
O_1\rightarrow O_6\rightarrow D_1\textrm{ or}\\
O_1\rightarrow D_2\rightarrow D_1\textrm{ or}\\
O_4\rightarrow D_2\rightarrow D_1,
\end{cases}\\
O_2&\rightarrow
\begin{cases}
O_1\rightarrow D_2\textrm{ or}\\
O_4\rightarrow D_2,
\end{cases}\\
O_2&\rightarrow
\begin{cases}
O_1\rightarrow D_2\rightarrow D_3\textrm{ or}\\
O_1\rightarrow D_4\rightarrow D_3\textrm{ or}\\
O_4\rightarrow D_2\rightarrow D_3,
\end{cases}\\
O_2&\rightarrow O_1\rightarrow D_4,\\
O_2&\rightarrow O_1\rightarrow D_5,\\
O_2&\rightarrow O_1\rightarrow D_6,\\
O_3&\rightarrow
\begin{cases}
O_1\rightarrow O_6\rightarrow D_1\textrm{ or}\\
O_1\rightarrow D_2\rightarrow D_1\textrm{ or}\\
O_4\rightarrow D_2\rightarrow D_1\textrm{ or}\\
O_5\rightarrow O_6\rightarrow D_1,
\end{cases}\\
O_3&\rightarrow
\begin{cases}
O_1\rightarrow D_2\textrm{ or}\\
O_4\rightarrow D_2,
\end{cases}\\
O_3&\rightarrow
\begin{cases}
O_1\rightarrow D_2\rightarrow D_3\textrm{ or}\\
O_1\rightarrow D_4\rightarrow D_3\textrm{ or}\\
O_4\rightarrow D_2\rightarrow D_3,
\end{cases}\\
O_3&\rightarrow O_1\rightarrow D_4,\\
O_3&\rightarrow O_1\rightarrow D_5,\\
O_3&\rightarrow O_1\rightarrow D_6,\\
O_4&\rightarrow D_2\rightarrow D_1,\\
O_4&\rightarrow D_2,\\
O_4&\rightarrow D_2\rightarrow D_3,\\
O_4&\rightarrow
\begin{cases}
O_1\rightarrow D_4\textrm{ or}\\
D_2\rightarrow D_4,
\end{cases}\\
O_4&\rightarrow
\begin{cases}
O_1\rightarrow D_5\textrm{ or}\\
D_2\rightarrow D_5,
\end{cases}\\
O_4&\rightarrow O_1\rightarrow D_6,\\
O_5&\rightarrow O_6\rightarrow D_1,\\
O_5&\rightarrow O_4\rightarrow D_2,\\
O_5&\rightarrow
\begin{cases}
O_4\rightarrow D_2\rightarrow D_3\textrm{ or}\\
O_6\rightarrow D_1\rightarrow D_3,
\end{cases}\\
O_5&\rightarrow
\begin{cases}
O_3\rightarrow O_1\rightarrow D_4\textrm{ or}\\
O_4\rightarrow O_1\rightarrow D_4\textrm{ or}\\
O_6\rightarrow O_1\rightarrow D_4\textrm{ or}\\
O_4\rightarrow D_2\rightarrow D_4,
\end{cases}\\
O_5&\rightarrow
\begin{cases}
O_3\rightarrow O_1\rightarrow D_5\textrm{ or}\\
O_4\rightarrow O_1\rightarrow D_5\textrm{ or}\\
O_6\rightarrow O_1\rightarrow D_5\textrm{ or}\\
O_4\rightarrow D_2\rightarrow D_5,
\end{cases}\\
O_5&\rightarrow
\begin{cases}
O_3\rightarrow O_1\rightarrow D_6\textrm{ or}\\
O_4\rightarrow O_1\rightarrow D_6\textrm{ or}\\
O_6\rightarrow O_1\rightarrow D_6,
\end{cases}\\
O_6&\rightarrow D_1,\\
O_6&\rightarrow
\begin{cases}
O_1\rightarrow D_2\textrm{ or}\\
D_1\rightarrow D_2,
\end{cases}\\
O_6&\rightarrow D_1\rightarrow D_3\\
O_6&\rightarrow O_1\rightarrow D_4\\
O_6&\rightarrow O_1\rightarrow D_5\\
O_6&\rightarrow O_1\rightarrow D_6.
\end{align}

The set of 24 ADE for the small-world network is:
\begin{gather}
O_1O_6,O_1D_2,O_1D_4,O_1D_5,O_1D_6,O_2O_4,O_2O_1,O_3O_4,\\
O_3O_5,O_3O_1,O_4D_2,O_4O_1,O_5O_6,O_5O_3,O_5O_4,O_6D_1,\\
O_6O_1,D_1D_2,D_1D_3,D_2D_3,D_2D_4,D_2D_5,D_2D_1,D_4D_3.
\end{gather}

Here are the edges that are being traversed by each origin node $O_i$ in the small-world network, to get to any destination:
\begin{align}
O_1&=O_1O_6,O_1D_2,O_1D_4,O_1D_5,O_1D_6,O_6D_1,D_2D_1,D_2D_3,\nonumber\\
&\phantom{=}D_4D_3\\
O_2&=O_2O_1,O_2O_4,O_1O_6,O_6D_1,O_1D_2,D_2D_1,O_4D_2,D_2D_3,\nonumber\\
&\phantom{=}O_1D_4,D_4D_3,O_1D_5,O_1D_6\\
O_3&=O_3O_1,O_3O_4,O_3O_5,O_1O_6,O_6D_1,O_1D_2,D_2D_1,O_4D_2,\nonumber\\
&\phantom{=}O_5O_6,D_2D_3,O_1D_4,D_4D_3,O_1D_5,O_1D_6\\
O_4&=O_4D_2,O_4O_1,D_2D_1,D_2D_3,O_1D_4,D_2D_4,O_1D_5,D_2D_5,\nonumber\\
&\phantom{=}O_1D_6\\
O_5&=O_5O_6,O_5O_4,O_5O_3,O_6D_1,O_4D_2,D_2D_3,D_1D_3,O_3O_1,\nonumber\\
&\phantom{=}O_1D_4,O_4O_1,O_6O_1,D_2D_4,O_1D_5,D_2D_5,O_1D_6\\
O_6&=O_6D_1,O_6O_1,O_1D_2,D_1D_2,D_1D_3,O_1D_4,O_1D_5,O_1D_6.
\end{align}

\begin{figure*}
\centering
\includegraphics[width=15cm]{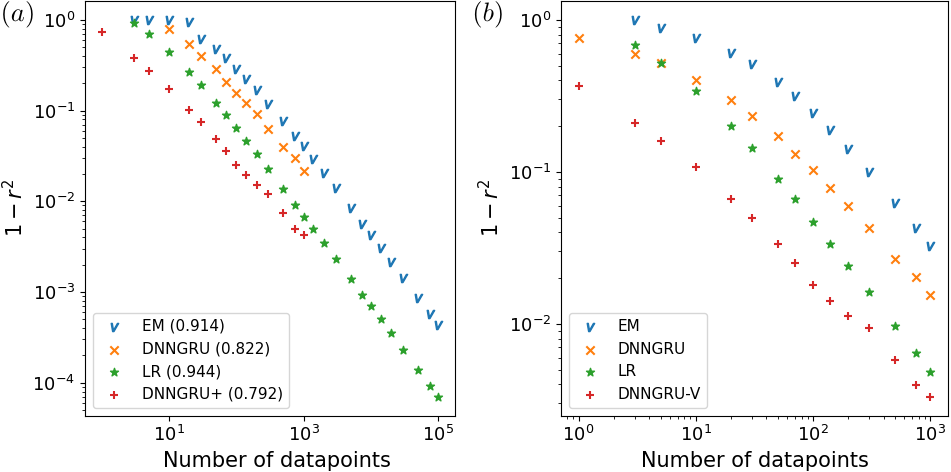}
\caption{a: The same graph as Fig.\ 3(c) in the main text, with more datapoints for EM and LR up till $10^5$. Shown in the parentheses of the legend are the corresponding exponents of the power law fits. The results for DNNGRU and DNNGRU-V are not produced beyond 1000 datapoints because they become computationally infeasible to train. Nevertheless, it is apparent by extrapolation that DNNGRU-V would converge to LR's performance (or even begin to become inferior) in the asymptotic limit with large datapoints. b: The corresponding results with respect to Fig.\ 3(c) in the main text, where the various methods are to predict the actual OD instead of $\zeta_{ij}$. Just like predicting $\zeta_{ij}$, LR, DNNGRU-V and DNNGRU are superior to Vardi's expectation-maximisation in predicting the actual OD.}
\label{figS1}
\end{figure*}

\begin{figure*}
\centering
\includegraphics[width=16cm]{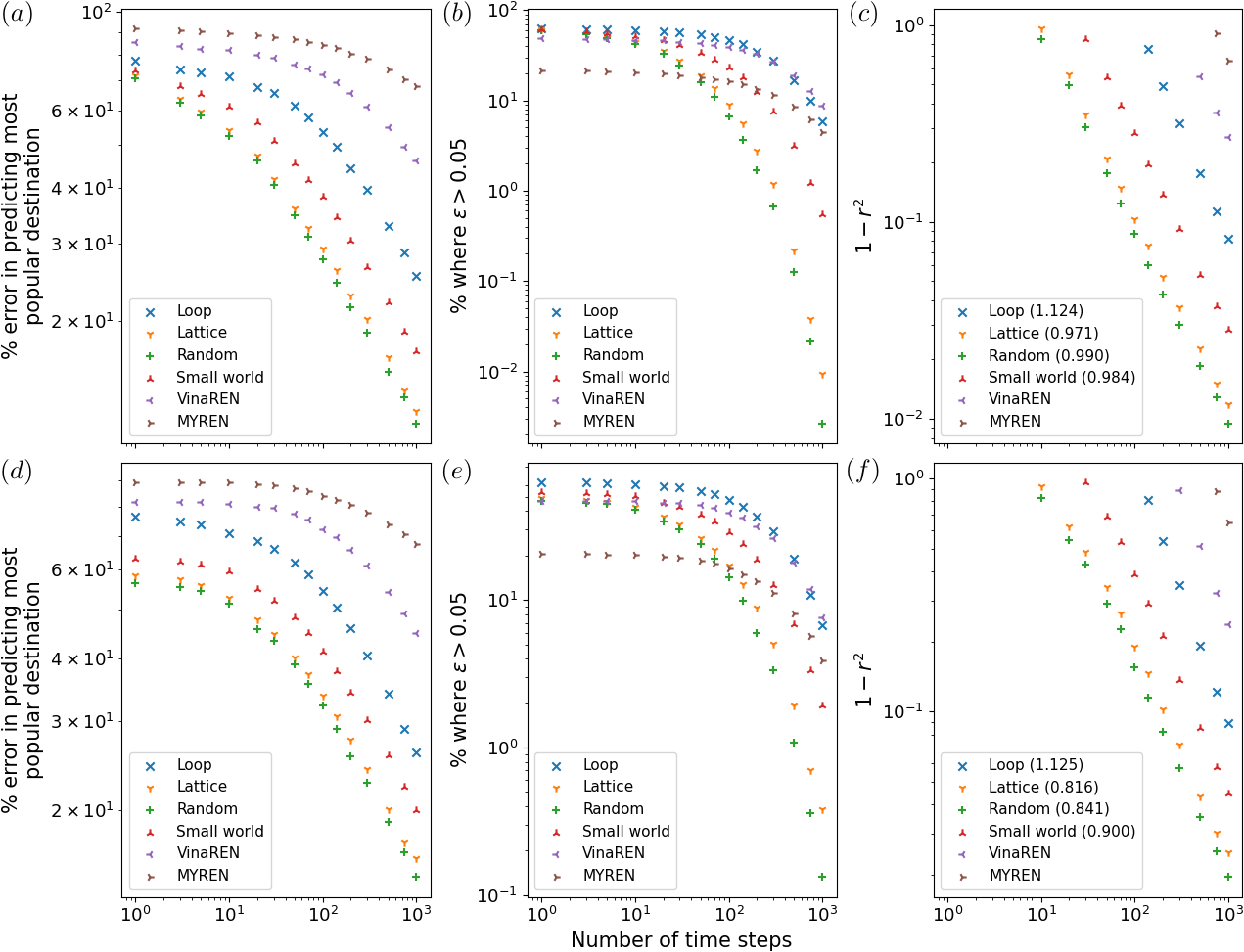}
\caption{The corresponding results with respect to Fig.\ 6 in the main text. The top row is for no lag ($l=1$) traversing the edges, whilst the bottom row is with lag of $l=10$ time steps.}
\label{figS2}
\end{figure*}

\begin{figure*}
\centering
\includegraphics[width=16cm]{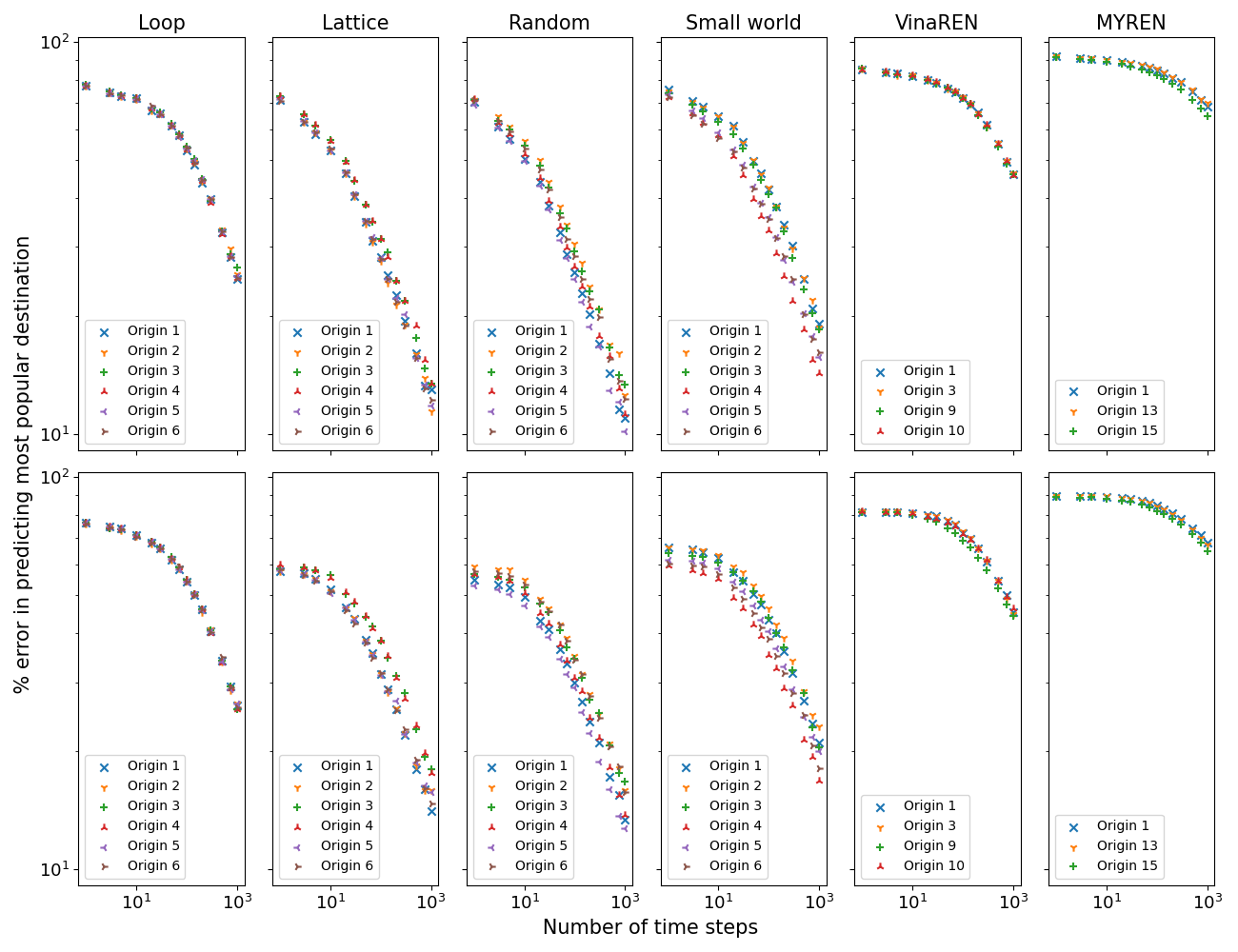}
\caption{The corresponding results with respect to Fig.\ 7 in the main text, for percentage error in predicting the most popular destination. The top row is for no lag ($l=1$) traversing the edges, whilst the bottom row is with lag of $l=10$ time steps.}
\label{figS3}
\end{figure*}

\begin{figure*}
\centering
\includegraphics[width=16cm]{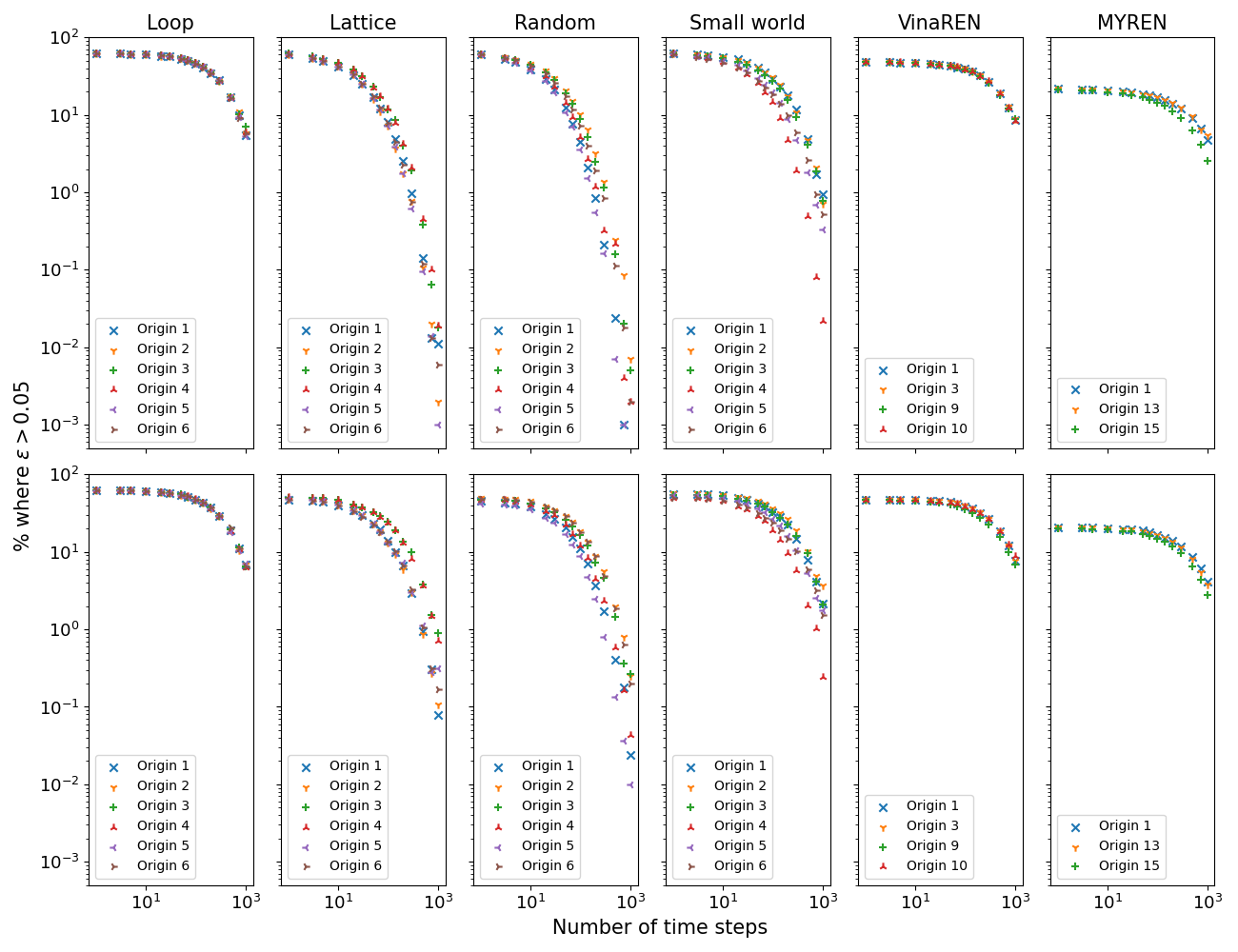}
\caption{The corresponding results with respect to Fig.\ 7 in the main text, for percentage of predictions where $\varepsilon>0.05$. The top row is for no lag ($l=1$) traversing the edges, whilst the bottom row is with lag of $l=10$ time steps.}
\label{figS4}
\end{figure*}

\begin{figure*}
\centering
\includegraphics[width=16cm]{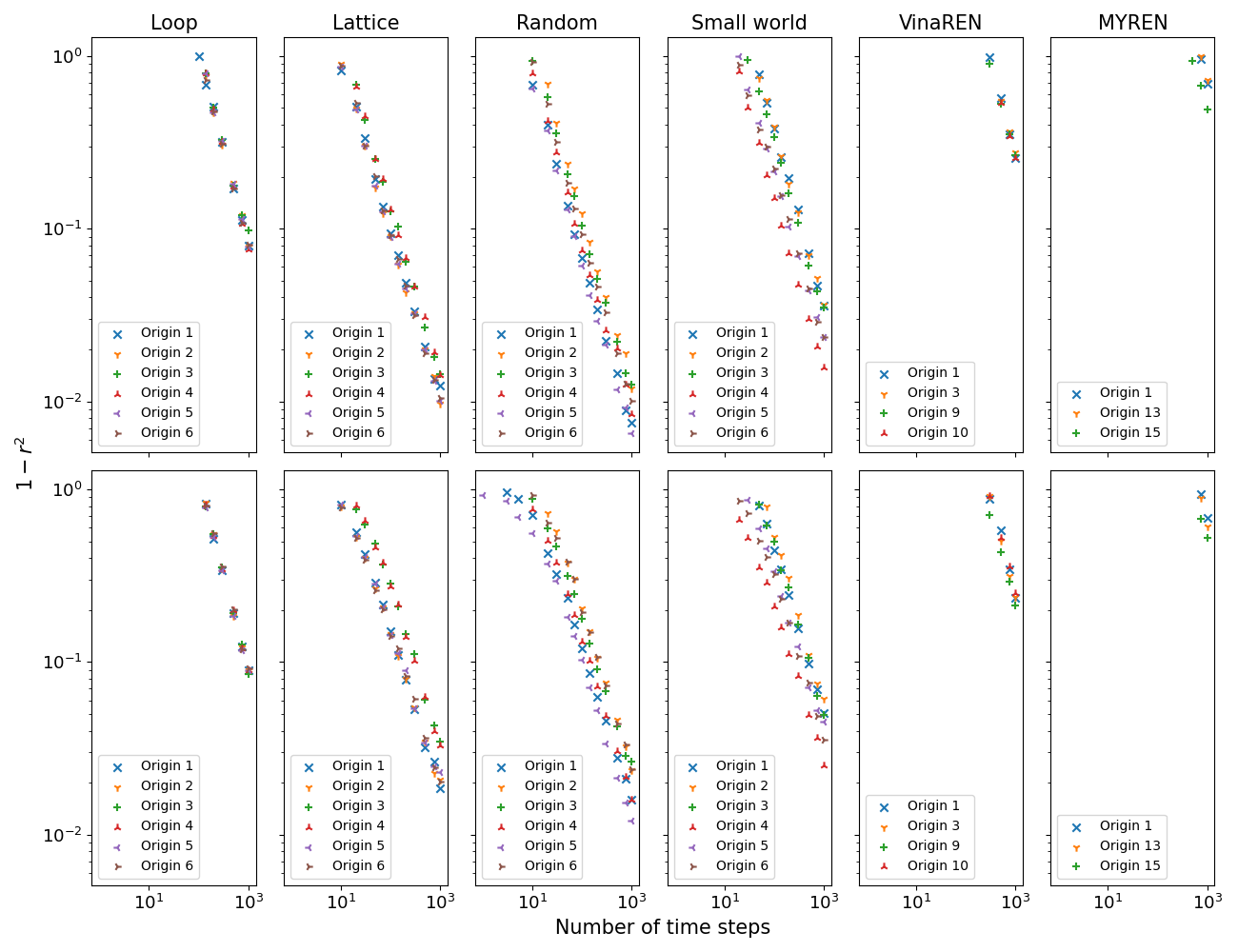}
\caption{The corresponding results with respect to Fig.\ 7 in the main text, for $1-r^2$. The top row is for no lag ($l=1$) traversing the edges, whilst the bottom row is with lag of $l=10$ time steps.}
\label{figS5}
\end{figure*}






\bibliography{Citation}





